\documentclass[a4paper,11pt]{article}
\pdfoutput=1

\usepackage{jheppub}
\usepackage{amsmath,amssymb,mathtools}
\usepackage{graphicx}
\usepackage{booktabs}
\usepackage{array,tabularx}
\usepackage{xcolor}
\usepackage{url}

\newcommand{\E}{\mathbb{E}}
\newcommand{\CN}{\mathcal{N}}
\newcommand{\one}{\mathbf{1}}

\title{\boldmath Virasoro Symmetry in Neural Network Field Theories}

\author{Brandon Robinson}

\affiliation{Institute of Physics, University of Amsterdam,
904 Science Park, 1098 XH Amsterdam, The Netherlands}

\emailAdd{b.j.robinson@uva.nl}

\abstract{
We construct the Virasoro algebra from a neural free-field architecture. A logarithmic momentum law and a smooth spectral window give the weight-one current kernel, whose radial-ordering discontinuity fixes the oscillator pairing. The limiting neural current correlations agree with the Fock vacuum correlations. Their Sugawara stress tensor gives the Virasoro algebra with central charge one, and charged vertex insertions identify its highest-weight sectors. Bounds on regulator and feature errors control finite-mode approximations to this representation. At finite width, an explicit adjoint-compatible projection defines an auxiliary Fock readout when its Hermitian mixed block is positive. Cauchy features extend the construction to fermionic and supersymmetric sectors, including a Clifford zero mode in the Ramond representation, while first-order stress tensors give fermionic and bosonic ghost systems. Shared reflected features impose scalar boundary conditions in every realization and yield conformal and superconformal gluing in the limiting fields. Numerical comparisons support the current pairing and field correlations; their detailed analysis is collected in the appendices.
}

\arxivnumber{2512.24420}

\begin{document}
\maketitle


\section{Introduction}
\label{sec:introduction}

We construct a realization of the Virasoro algebra from a neural free-field architecture. The choice of Fourier features and their parameter measure fixes a chiral current kernel. Its radial-ordering discontinuity determines the oscillator pairing, whose Fock quantization gives a stress tensor and its centrally extended mode algebra. This connects a specified neural ensemble to the generators of local conformal transformations in two dimensions \cite{Belavin:1984vu,DiFrancesco:1997nk,GoddardOlive:1986}.

Within the broader interface between physics and machine learning \cite{Carleo:2019ptp}, wide neural networks provide a general route to Gaussian fields \cite{Neal:1996,NIPS1996_ae5e3ce4,Lee:2017wiy,Halverson:2020trp}. Random cosine features realize free propagators through their spectral weights \cite{Halverson:2021aot}, while embedding-space constructions produce conformal fields and stress-tensor data \cite{Halverson:2024axc}. Symmetry constraints underlie geometric deep learning \cite{Bronstein:2021} and invariant random-feature and kernel models \cite{Mei:2021colt}; at the ensemble level they are encoded in the parameter distribution \cite{Maiti:2021fpy}. Finite-width cumulants quantify departures from Gaussianity \cite{Dyer:2019uzd,Yaida:2020,Roberts:2021principles,Demirtas:2023fir,Grosvenor:2022edge}. Here the architecture is used to construct the chiral modes and their Virasoro action. The limiting neural current moments agree with the radially ordered Fock vacuum moments, and the central coefficient follows from the resulting oscillator contractions. The contour feature map, its limiting pairing, and its action on charged level sectors are explicit parts of the construction.

For the scalar architecture, the weight-one current selects $p(k)\propto|k|^{-2}$ within the isotropic power-law family. A smooth spectral window controls the differentiated kernel on separated contours. The two radial orderings then give $\omega_{mn}=m\delta_{m+n,0}$ in the continuum limit. Canonical quantization of this form yields the oscillator representation, and the Sugawara construction gives the Virasoro algebra with $c=1$. We determine the oscillator support required for its exact action on each fixed charged level sector. Bounds on the regulated and sampled pairings control the approach to this action.

The quantum representation enters through Fock quantization of the limiting pairing; the neural fields themselves are commuting random variables. At finite width, an empirical pairing need not satisfy the adjoint and positivity conditions of that representation. We give an explicit mixed-block projection that defines a finite-window oscillator readout when its Hermitian block is positive. The projection correction, positivity probability, and convergence of finite-level matrix elements are controlled by the original pairing error. Covariances and charged-field Ward ratios are evaluated on the original neural ensemble.

Neural worldsheet constructions and string amplitudes have also been developed in \cite{Frank:2026bui}. The present work resolves the chiral mode algebra and its central extension directly from the feature ensemble, providing the local conformal structure for such applications.

Using the Grassmann neural-field formalism of~\cite{Frank:2025fermions}, we construct Majorana and Dirac sectors from Cauchy features and obtain the free supersymmetric completion. An explicit Clifford zero mode extends the construction to the Ramond representation. First-order stress tensors give the $bc$ and $\beta\gamma$ systems, with their conformal weights and central charges. Shared reflected features impose Dirichlet or Neumann conditions in every realization and give conformal and superconformal boundary gluing in the limiting fields. This complements embedding-space treatments of conformal defects \cite{Billo:2016cpy,Capuozzo:2025ozt}.

Section~\ref{sec:architectures} specifies the architectures and parameter spaces. Section~\ref{sec:pairing} carries the scalar construction through to the Virasoro algebra, and section~\ref{sec:readouts} develops its charged, fermionic, and boundary sectors. The numerical comparisons in section~\ref{sec:audits} support the analytical construction; their full results, methods, and uncertainties are collected in the numerical appendices. The kernel estimates and finite-mode arguments remain in the appendices. 

We use the smooth spectral-regulator and ordered-limit framework developed with Pietro Capuozzo and Benjamin Suzzoni and used in the WZW constructions of~\cite{CapuozzoRobinsonSuzzoni:2026}. Here we specialize it to constant-amplitude Fourier features with a raised-cosine window, derive the current and pairing estimates, and connect them to the Virasoro representation.

\section{Neural architectures and parameter ensembles}
\label{sec:architectures}
\label{sec:lk}

We use random-feature fields whose correlations are fixed by a parameter functional.  The scalar parameters are momenta and phases, $\Theta_B=\{k_j,\gamma_j\}_{j=1}^N$, with probability measure
\begin{equation}
d\mu_B(\Theta_B)=\prod_{j=1}^N  p_{\Lambda,\tau}(k_j)\,d^2k_j\,\frac{d\gamma_j}{2\pi}.
\label{eq:bosonic-parameter-space}
\end{equation}
The density is normalized at finite cutoffs. For fermionic fields we also integrate over Berezin coefficients; a Cauchy representation below samples discrete mode indices. Spatial averaging and coefficient contraction will be indicated explicitly. Sections~\ref{sec:pairing} and~\ref{sec:readouts} develop the operator readouts of these fields.

\subsection{Scalar fields and the logarithmic prior}

Let $x=(x^1,x^2)\in\mathbb R^2$, $z=x^1+i x^2$, and
\begin{equation}
\phi_N(x;\Theta) = \frac{\sigma_\phi}{\sqrt N} \sum_{j=1}^{N}\cos(k_j\cdot x+\gamma_j), \qquad \gamma_j\sim {\rm Unif}[0,2\pi).
\label{eq:lk-field}
\end{equation}
Define the spectral-cutoff pair
\begin{equation}
\Lambda_{\rm spec}:=(\Lambda_{\rm IR},\Lambda_{\rm UV})
\end{equation}
and write
\begin{equation}
u(k)= \frac{\log(k/\Lambda_{\rm IR})} {\log(\Lambda_{\rm UV}/\Lambda_{\rm IR})}.
\end{equation}
Within the smooth-regulator framework of~\cite{CapuozzoRobinsonSuzzoni:2026}, we choose a raised-cosine window in $\log k$, 
\begin{equation}
W_\tau(u)= \begin{cases} \frac12\left[1-\cos\!\left(\pi u/\tau\right)\right], &0\le u<\tau,\\
1,&\tau\le u\le1-\tau,\\
\frac12\left[1-\cos\!\left(\pi(1-u)/\tau\right)\right], &1-\tau<u\le1,\\
0,&\text{otherwise},
\end{cases}
\label{eq:raised-cosine-window}
\end{equation}
with $\tau=0.15$.  The momentum density is
\begin{equation}
p_{\Lambda,\tau}(k) = \frac{1}{\mathcal Z_{\Lambda,\tau}} \frac{W_\tau(u(|k|))}{|k|^2}, \qquad \mathcal Z_{\Lambda,\tau} = 2\pi\int_{\Lambda_{\rm IR}}^{\Lambda_{\rm UV}} \frac{dk}{k}\,W_\tau(u(k)).
\label{eq:lk-prior}
\end{equation}
The limit $\tau=0$ denotes the sharp annular window $\one_{[\Lambda_{\rm IR},\Lambda_{\rm UV}]}$ and is retained throughout as a regulator control. The taper changes the two endpoint regions and retains the interior law $|k|^{-2}$.

The spectral covariance depends on the product of the momentum density and the squared feature amplitude. Here the massless spectral weight is placed in the density and the cosine features have constant amplitude. Other factorizations give the same Gaussian covariance while changing higher moments and sampling variance \cite{Halverson:2021aot}.

Phase averaging gives the translation- and rotation-invariant covariance
\begin{align}
K_{\Lambda,\tau}(x-y) &:=\mathbb E_\Theta[\phi_N(x)\phi_N(y)] \\
&= \frac{\sigma_\phi^2}{2} \int d^2k\,p_{\Lambda,\tau}(k) \cos(k\cdot(x-y)). 
\label{eq:finite-kernel}
\end{align}
This covariance is already independent of $N$; infinite width makes the field Gaussian.  For
\begin{equation}
\Lambda_{\rm UV}^{-1}\ll |x-y|\ll\Lambda_{\rm IR}^{-1},
\label{eq:physical-window}
\end{equation}
the scalar kernel has the logarithmic form
\begin{equation}
K_{\Lambda,\tau}(x-y) =-\kappa\log\!\bigl(\mu^2|x-y|^2\bigr) +C_{\Lambda,\tau}+\delta K_{\Lambda,\tau}(x-y),
\label{eq:log-kernel}
\end{equation}
with the normalization fixed to $\kappa=1$.  Because $p_{\Lambda,\tau}$ is normalized as a probability density, this requires the feature amplitude to scale with the logarithmic spectral volume.  Writing
\begin{equation}
\mathcal Z_{\Lambda,\tau} =2\pi\int_{\Lambda_{\rm IR}}^{\Lambda_{\rm UV}} \frac{dk}{k}\,W_\tau(u(k)),
\end{equation}
the convention $K(r)=-\log(\mu^2r^2)+\cdots$ fixes
\begin{equation}
{\sigma_\phi^2=\frac{2\mathcal Z_{\Lambda,\tau}}{\pi}}.
\label{eq:field-normalization}
\end{equation}
For the precise window in \eqref{eq:raised-cosine-window}, $\int_0^1 W_\tau(u)\,du=1-\tau$, so the normalization is also
\begin{equation}
\mathcal Z_{\Lambda,\tau}=2\pi(1-\tau) \log\frac{\Lambda_{\rm UV}}{\Lambda_{\rm IR}},\qquad \sigma_\phi^2=4(1-\tau) \log\frac{\Lambda_{\rm UV}}{\Lambda_{\rm IR}}. 
\label{eq:exact-taper-mass}
\end{equation}
In numerical calculations we use $\sigma_\phi^2=4L_w$, with $L_w$ evaluated by quadrature. Equation~\eqref{eq:log-kernel} keeps the infrared constant and cutoff remainder explicit over the scaling window. 

The four-point function makes the approach to the Gaussian field
explicit. Let $X_i=\sigma_\phi\cos(k\cdot x_i+\gamma)$ denote
one feature evaluated at $x_i$, and set
$K_{ij}=K_{\Lambda,\tau}(x_i-x_j)$ and
$W_4=K_{12}K_{34}+K_{13}K_{24}+K_{14}K_{23}$.
Independence of the Fourier features gives the exact relation
\begin{equation}
\begin{aligned}
G_{4,N}(x_1,x_2,x_3,x_4)
&:=\E_\Theta\!\left[\prod_{i=1}^4\phi_N(x_i)\right]\\
&=W_4+\frac1N C_4^{(1)}(x_1,x_2,x_3,x_4),\\
G_{4,N}^{\rm conn}
&:=G_{4,N}-W_4=\frac1N C_4^{(1)},
\end{aligned}
\label{eq:four-point-width-scaling}
\end{equation}
where $C_4^{(1)}=\E[X_1X_2X_3X_4]-W_4$.
At fixed spectral regulator this coefficient is independent of
$N$. The connected four-point function therefore decays as
$1/N$, and the full four-point function tends to its Wick
contractions. The feature contraction and its equal-point limit
are derived in appendix~\ref{appnum-app:four-point-width}.

Differentiating the scalar field defines the current.  We use
\begin{equation}
\partial=\frac12(\partial_{x^1}-i\partial_{x^2}), \qquad J_N(z;\Theta)=i\,\partial\phi_N(z;\Theta).
\label{eq:current-definition}
\end{equation}
For $\mathbb E[\phi(z,\bar z)\phi(w,\bar w)]= -\log|z-w|^2$, the target holomorphic kernel is
\begin{equation}
K_{JJ}(z,w)=\mathbb E[J(z)J(w)]=\frac{1}{(z-w)^2}.
\label{eq:target-current-kernel}
\end{equation}
At finite cutoff the exact kernel is obtained by differentiating $K_{\Lambda,\tau}$,
\begin{equation}
K^{\Lambda,\tau}_{JJ}(z,w) =-\partial_z\partial_w K_{\Lambda,\tau}(z-w).
\label{eq:regulated-current-kernel}
\end{equation}
This kernel is smooth below the UV resolution scale and generally non-holomorphic. We retain both contour radii when extracting its mode pairing. Section~\ref{sec:audits} compares its dependence on the sharp and tapered spectral endpoints.

We determine the spectral exponent within the rotationally invariant power-law family
\begin{equation}
p_\alpha(k)\propto |k|^{-\alpha}
\end{equation}
using the holomorphic derivative $\partial=\tfrac12(\partial_{x^1}-i\partial_{x^2})$. If $x=re^{i\theta_x}$, each same-chirality derivative contributes one unit of angular spin, and the angular integral is
\begin{equation}
\int_0^{2\pi}d\theta_k\,  e^{ikr\cos(\theta_k-\theta_x)}e^{-2i\theta_k} =-2\pi e^{-2i\theta_x}J_2(kr).
\label{eq:spin-two-angular-integral}
\end{equation}
At finite cutoffs the same-chirality derivative covariance has the well-defined spin-two transform
\begin{equation}
G_{\partial\phi}^{\Lambda}(x) \propto e^{-2i\theta_x} \int_{\Lambda_{\rm IR}}^{\Lambda_{\rm UV}}  dk\,k^{3-\alpha}W_\tau(u(k))J_2(kr). 
\label{eq:powerlaw-current}
\end{equation}
Rescaling $q=kr$ factors out $r^{-(4-\alpha)}$ when the regulated Bessel integral has a nonzero scaling-window limit. At $\alpha=2$, the improper integral $\int_0^\infty qJ_2(q)\,dq$ diverges. Abel regularization with $e^{-\epsilon q}$ gives the limit $2$ as $\epsilon\downarrow0$. We use the smooth $W_\tau$ above; its regulator-removal estimate is derived in appendix~\ref{app:continuum-estimates}. A dimension-one holomorphic current requires $\langle J(z)J(0)\rangle\propto z^{-2} =e^{-2i\theta_x}r^{-2}$, and therefore \begin{equation} {\alpha=2}. \end{equation} The logarithmic spectrum therefore has the required current scaling.

\subsection{Fermionic and first-order fields}
\label{sec:fermion-architectures}
\label{sec:spin-architectures}
\label{sec:first-order-architectures}

For each feature introduce a Berezin pair $(\xi_j,\bar\xi_j)$ with
\begin{equation}
d\mu_G=\prod_{j=1}^N d\bar\xi_j\,d\xi_j\, \exp\!\left(\sum_j\xi_j\bar\xi_j\right),\qquad \langle1\rangle_G=1,\qquad \langle\xi_j\bar\xi_l\rangle_G=\delta_{jl}.
\label{eq:berezin-parameter-functional}
\end{equation}
The integration order is chosen so that $\int d\bar\xi\,d\xi\,\xi\bar\xi=1$; reversing the two coefficients therefore reverses the contraction sign.  This defines an algebraic parameter functional on the Grassmann algebra. Grassmann neural fields and their Wick contractions are discussed in  \cite{Frank:2025fermions}. The continuum fermionic kernel is
\begin{equation}
K_{\psi\psi}(z,w)=\frac{1}{z-w}.
\end{equation}
For $|z|>|w|$ its mode expansion is
\begin{equation}
\frac1{z-w} = \sum_{n\ge0}z^{-n-1}w^n,
\end{equation}
which gives an independent separable feature representation.  Draw independent feature indices $n_j\in\mathbb Z_{\ge0}$ with $\mathbb P(n_j=n)=q_n>0$ and introduce Berezin pairs $(\xi_j,\bar\xi_j)$ with $\langle\xi_j\bar\xi_l\rangle_G=\delta_{jl}$ and $\langle\bar\xi_l\xi_j\rangle_G=-\delta_{jl}$.  On the radially ordered domains set
\begin{equation}
\psi_{+,N}(z)=\frac1{\sqrt N}\sum_{j=1}^N \frac{\xi_j z^{-n_j-1}}{\sqrt{q_{n_j}}},\qquad \psi_{-,N}(w)=\frac1{\sqrt N}\sum_{j=1}^N \frac{\bar\xi_j w^{n_j}}{\sqrt{q_{n_j}}}.
\label{eq:random-cauchy-fermions}
\end{equation}
At fixed sampled indices, its Berezin contraction is the finite sum
\begin{equation}
K_{\psi\psi,N}^{>}(z,w)=\frac1N\sum_{j=1}^N \frac{z^{-n_j-1}w^{n_j}}{q_{n_j}},\qquad \mathbb E_n K_{\psi\psi,N}^{>}(z,w)=\frac1{z-w} \quad(|w|<|z|).
\label{eq:finite-cauchy-kernel}
\end{equation}
The opposite radial ordering uses the exchanged Grassmann contraction and its minus sign.  In the deterministic alternative one truncates the geometric series at $n=L-1$, giving $K_L^{>}(z,w)=[1-(w/z)^L]/(z-w)$.

The momentum construction can also be written directly as a neural field. Set $k_j=\kappa_j e^{i\vartheta_j}$, $s_j(z)=\operatorname{Re}(\bar k_j z)$, and 
\begin{equation}
A_j=\sqrt{L_w\kappa_j}\,e^{-i\vartheta_j/2},\qquad u_j(z)=A_j\cos s_j(z),\qquad v_j(z)=-A_j\sin s_j(z).
\label{eq:native-majorana-basis}
\end{equation}
With the same normalized momentum law as the scalar, define
\begin{equation}
\psi_N^{\rm native}(z)=\frac1{\sqrt N}\sum_{j=1}^N \bigl[\xi_j u_j(z)+\bar\xi_j v_j(z)\bigr].
\label{eq:native-majorana-field}
\end{equation}
The two basis factors carry half of the angular phase each, so their antisymmetric contraction has spin one.  At fixed momenta, $\langle\psi_N(z)\psi_N(w)\rangle_G =N^{-1}\sum_j[u_j(z)v_j(w)-v_j(z)u_j(w)]$.

The resulting finite-feature kernel is
\begin{equation}
K_{\psi\psi,N}^{\rm native}(z,w) =\frac{L_w}{N}\sum_{j=1}^{N}\overline{k_j} \sin\!\left(\operatorname{Re}[\overline{k_j}(z-w)]\right),
\label{eq:native-ns-kernel}
\end{equation}
with its ordered contour pairing evaluated before any CAR canonicalization. It is obtained by differentiating the scalar covariance and remains non-holomorphic at finite cutoff. The independent Cauchy representation uses its own mode draws. Appendix~\ref{app:cauchy-sampling} gives the proposal conditions needed for finite variance and retained-mode sampling. 

The mode functions also encode the choice of spin structure.  The Cauchy series above uses single-valued powers on the plane and hence half-integer NS oscillator labels.  If $z=e^u$, a weight-$1/2$ field on the cylinder is $\psi_{\rm cyl}(u)=z^{1/2}\psi(z)$: a single-valued plane field is antiperiodic on the cylinder.  A Ramond field instead has
\begin{equation}
\psi^{\rm R}(z)=\sum_{n\in\mathbb Z}b_n z^{-n-1/2}, \qquad \{b_n,b_m\}=\delta_{n+m,0},
\label{eq:ramond-mode-field}
\end{equation}
and is periodic on the cylinder.  Its nonzero modes admit the same importance-sampled construction: draw $m_j\ge1$ with probabilities $\varpi_m>0$ and use the ordered features $z^{-m_j-1/2}/\sqrt{\varpi_{m_j}}$ and $w^{m_j-1/2}/\sqrt{\varpi_{m_j}}$ with a Berezin pair.  After adjoining the zero-mode readout described in section~\ref{sec:ramond-readout}, their ordered contraction is
\begin{equation}
S_{{\rm R},N}^{>}(z,w)=\frac1{\sqrt{zw}} \left[\frac12+\frac1N\sum_{j=1}^N \frac{(w/z)^{m_j}}{\varpi_{m_j}}\right],\qquad |w|<|z|.
\label{eq:ramond-feature-kernel}
\end{equation}
The square roots use one consistent branch.  Averaging the mode indices gives the Ramond ground-state kernel
\begin{equation}
S_{\rm R}^{>}(z,w)=\frac1{\sqrt{zw}} \left[\frac12+\sum_{m\ge1}(w/z)^m\right] =\frac{z+w}{2\sqrt{zw}(z-w)}.
\label{eq:ramond-ground-kernel}
\end{equation}
Its short-distance singularity is still $1/(z-w)$, while the branch structure and zero-mode contribution differ from the NS vacuum.  A change of spin structure therefore changes both the mode functions and the state in which they are contracted.

A complex field follows by combining two independent Majorana ensembles,
\begin{equation}
\Psi=\frac{\psi_1+i\psi_2}{\sqrt2},\qquad \Psi^\dagger=\frac{\psi_1-i\psi_2}{\sqrt2}.
\label{eq:dirac-architecture}
\end{equation}
The limiting cross contraction of this pair is $1/(z-w)$, and contractions of two $\Psi$ fields or two $\Psi^\dagger$ fields vanish.  The associated current and its relation to the scalar are given in section~\ref{sec:dirac-readout}.

The same separable Cauchy features supply first-order pairs $(F,G)$. Write $f_n(z)=z^{-n-1}$ and $g_n(z)=z^n$, draw $n_j$ with the actual proposal probabilities $q_n$, and take two independent coefficient pairs $(a_j,b_j)$ and $(c_j,d_j)$.  Define the ordered pieces
\begin{align}
F_{+,N}(z)&=\frac{\kappa}{\sqrt N}\sum_j \frac{a_j f_{n_j}(z)}{\sqrt{q_{n_j}}},& G_{-,N}(z)&=\frac1{\sqrt N}\sum_j \frac{b_j g_{n_j}(z)}{\sqrt{q_{n_j}}},\nonumber\\
G_{+,N}(z)&=\frac1{\sqrt N}\sum_j \frac{c_j f_{n_j}(z)}{\sqrt{q_{n_j}}},& F_{-,N}(z)&=\frac1{\sqrt N}\sum_j \frac{d_j g_{n_j}(z)}{\sqrt{q_{n_j}}}.
\label{eq:first-order-features}
\end{align}
For anticommuting $bc$ fields, take $\kappa=+1$ and use two Berezin pairs.  For commuting $\beta\gamma$ fields, take $\kappa=-1$, $b_j=\bar a_j$, and $d_j=\bar c_j$, where $a_j,c_j$ are independent standard circular complex Gaussians.  In both cases the basic pair contractions are $\langle a_jb_l\rangle=\langle c_jd_l\rangle=\delta_{jl}$. At fixed mode indices this gives
\begin{equation}
\langle F_{+,N}(z)G_{-,N}(w)\rangle=\kappa C_N^>(z,w),\qquad \langle G_{+,N}(z)F_{-,N}(w)\rangle=C_N^>(z,w),
\label{eq:first-order-kernels}
\end{equation}
where $C_N^>$ is the finite Cauchy sum in \eqref{eq:finite-cauchy-kernel}; equal-field contractions vanish. The full local fields are defined by the first-order oscillator relations with these ordered contractions.  The readout parameter $\lambda$ assigns weights $(\lambda,1-\lambda)$ through the stress tensor in section~\ref{sec:ghost-readouts}.

\subsection{Reflected fields on the upper half-plane}
\label{sec:boundary-architectures}

For $\mathbb H=\{z:\operatorname{Im}z>0\}$, reflection imposes the boundary conditions directly on the neural features.  This is complementary to the embedding-space defect construction of \cite{Capuozzo:2025ozt}. Using shared momenta and phases at a point and its image, define
\begin{equation}
\Phi_{\sigma,N}(z,\bar z) =\frac{\phi_N(z,\bar z)+\sigma\phi_N(\bar z,z)}{\sqrt2}, \qquad \sigma\in\{+1,-1\}.
\label{eq:boundary-scalar-features}
\end{equation}
Here $\phi_N(\bar z,z)$ means evaluation of the real Fourier field at $(x,-y)$.  Reflection makes $\Phi_{+,N}$ even and $\Phi_{-,N}$ odd in $y$, so every realization satisfies
\begin{equation}
\left.\partial_y\Phi_{+,N}\right|_{y=0}=0, \qquad \left.\Phi_{-,N}\right|_{y=0}=0.
\label{eq:boundary-exact-bcs}
\end{equation}
These are respectively Neumann and homogeneous Dirichlet conditions. The factor $1/\sqrt2$ preserves the normalization of the direct bulk singularity.

The isotropic ensemble covariance obeys the image formula at every width and finite spectral cutoff:
\begin{equation}
K^\sigma_{\Lambda,\tau}(z,w) =K_{\Lambda,\tau}(z-w) +\sigma K_{\Lambda,\tau}(z-\bar w).
\label{eq:boundary-regulated-kernel}
\end{equation}
This follows by expanding the four terms in $\mathbb E[\Phi_{\sigma,N}(z)\Phi_{\sigma,N}(w)]$ and using reflection invariance of the momentum law.  At a fixed momentum sample the conditional covariance retains all four terms. Ensemble reflection invariance combines them into the two terms above.  On separated points in the scaling window, the limiting image kernel is
\begin{equation}
K^\sigma(z,w) =-\log|z-w|^2-\sigma\log|z-\bar w|^2,
\label{eq:boundary-log-kernel}
\end{equation}
modulo the scalar infrared constant.  This constant cancels for Dirichlet conditions and is removed by taking covariance differences for Neumann conditions.  At fixed regulator, infinite width gives a Gaussian field with covariance \eqref{eq:boundary-regulated-kernel}. Subsequent regulator removal yields the conformal image kernel.

Let $\chi$ be the doubled chiral Majorana field whose CAR readout is described in section~\ref{sec:fermionic-modes}.  A reflecting boundary is specified by
\begin{equation}
\psi_L(z)=\chi(z),\qquad \psi_R(\bar z)=\eta\chi(\bar z),\qquad \eta=\pm1.
\label{eq:boundary-doubled-fermion}
\end{equation}
Thus $\psi_L(x)=\eta\psi_R(x)$ on the real axis.  At finite feature width this relation can be imposed separately on the positive- and negative-frequency features in \eqref{eq:random-cauchy-fermions}: the right sector evaluates the same mode indices and Berezin pairs at the reflected argument, with the factor $\eta$.  The reflection relation is then exact for each ordered feature field.  The kernel limit and retained CAR modes satisfy the convergence and sampling conditions of section~\ref{sec:fermionic-modes} and appendix~\ref{app:cauchy-sampling}.

\section{From neural features to the Virasoro algebra}
\label{sec:pairing}

For each parameter realization, $\phi_N$ and the projected coefficients built from it are commuting random variables.  In particular,
\begin{equation}
    [a_m(\Theta),a_n(\Theta)]_{\rm pointwise}=0.
\end{equation}
The discontinuity between two radial orderings of the current kernel supplies the central form. We construct the quantum representation from its limit and use finite samples to estimate the chiral data.


\subsection{Ordered contour modes}

Choose a circle $|z|=R$ lying inside the physical window.  The projected current coordinates are
\begin{equation}
a_m^{(N)}(R;\Theta) = \oint_{|z|=R}\frac{dz}{2\pi i}\,z^m J_N(z;\Theta), \qquad m\in\mathbb Z.
\label{eq:classical-mode-projection}
\end{equation}
The Jacobi--Anger expansion gives the Bessel-weighted finite feature map recorded in appendix~\ref{app:mode-projection}.  The nonzero modes extract the chiral pairing; the zero mode is specified in section~\ref{sec:quantized}.

For a finite mode window with oscillator cutoff $L$, we use
\begin{equation}
\mathcal M_L^\times = \{-L,\ldots,-1,1,\ldots,L\}.
\label{eq:nonzero-mode-window}
\end{equation}
The full charged Fock construction also contains $\alpha_0$.

Let $K_{JJ}(z,w)$ be a chiral current kernel in an annulus contained in the scaling window, and take $R_{\rm out}>R_{\rm in}>0$.  Define  
\begin{align}
\mathcal I^{>}_{mn}[K] &= \oint_{|z|=R_{\rm out}}\frac{dz}{2\pi i}\,z^m \oint_{|w|=R_{\rm in}}\frac{dw}{2\pi i}\,w^n K(z,w),
\label{eq:pairing-order-greater}
\\
\mathcal I^{<}_{mn}[K] &= \oint_{|z|=R_{\rm in}}\frac{dz}{2\pi i}\,z^m \oint_{|w|=R_{\rm out}}\frac{dw}{2\pi i}\,w^n K(z,w).
\label{eq:pairing-order-less}
\end{align}
The induced bilinear form is the ordering discontinuity
\begin{equation}
\omega_{mn}[K] = \mathcal I^{>}_{mn}[K]-\mathcal I^{<}_{mn}[K].
\label{eq:ordered-discontinuity-pairing}
\end{equation}
Equation~\eqref{eq:ordered-discontinuity-pairing} retains the radial-ordering information in the kernel.

At finite width the same contour functional acts on the empirical kernel.  For a fixed momentum sample $\{k_j\}_{j=1}^{N}$, let
\begin{equation}
K_{JJ}^{(N,\Lambda,\tau)}(z,w) = \mathbb E_{\gamma}\!\left[ J_N(z;\{k_j,\gamma_j\})J_N(w;\{k_j,\gamma_j\}) \,\middle|\,\{k_j\} \right]
\label{eq:finite-feature-current-kernel}
\end{equation}
be the phase-averaged empirical current kernel.  Equivalently, the exact regulated spectral measure is replaced by the empirical measure $N^{-1}\sum_j\delta_{k_j}$.  The architecture-level pairing is
\begin{equation}
\omega_{mn}^{(N,\Lambda,\tau)}(R_{\rm in},R_{\rm out}) = \mathcal I^{>}_{mn}[K_{JJ}^{(N,\Lambda,\tau)}] - \mathcal I^{<}_{mn}[K_{JJ}^{(N,\Lambda,\tau)}].
\label{eq:operational-finite-pairing}
\end{equation}
We express the finite pairing in dimensionless $R_0$ units.  Under $\zeta=z/R_0$ a chiral current of weight one transforms as $J_\zeta(\zeta)=R_0J_z(R_0\zeta)$; hence its mode coefficients and ordered pairing obey
\begin{equation}
 a_m^{(\zeta)}=R_0^{-m}a_m^{(z)},\qquad  \widetilde\omega_{mn}=R_0^{-(m+n)}\omega_{mn}.
 \label{eq:r0-unit-pairing}
\end{equation}
This preserves the canonical target because its nonzero entries have $m+n=0$.  In subsequent finite-feature matrix estimates $\Omega$ denotes $\widetilde\omega$ in this normalized contour coordinate; the unnormalized contour functional retains the symbol $\omega$. For each regulator, choose an evaluation interval $[r_{\min},r_{\max}]$ inside \eqref{eq:physical-window} and set
\begin{equation}
R_0=\sqrt{r_{\min}r_{\max}},\qquad R_{\rm in}=\frac{R_0}{\sqrt\rho},\qquad R_{\rm out}=R_0\sqrt\rho .
\label{eq:operational-radii}
\end{equation}
The reference calculation uses $\rho=2$.  To examine radius dependence, we vary $R_0$ and $\rho$ while maintaining $r_{\min}<R_{\rm in}<R_{\rm out}<r_{\max}$.

A covariance projected on one circle is symmetric, so its antisymmetrization vanishes. The discontinuity in \eqref{eq:operational-finite-pairing} compares projections on nested circles and retains the radial ordering.

For $K_{JJ}(z,w)=(z-w)^{-2}$,
\begin{equation}
\omega_{mn}=m\delta_{m+n,0}.
\label{eq:canonical-pairing}
\end{equation}
For $|z|>|w|$,
\begin{equation}
\frac{1}{(z-w)^2} = \sum_{k\geq0}(k+1)w^k z^{-k-2},
\end{equation}
so the two residues require $k=m-1=-n-1$ and give
\begin{equation}
\mathcal I^{>}_{mn} = m\,\mathbf 1_{m>0}\delta_{m+n,0}.
\end{equation}
For $|w|>|z|$,
\begin{equation}
\frac{1}{(z-w)^2} = \sum_{k\geq0}(k+1)z^k w^{-k-2},
\end{equation}
which gives
\begin{equation}
\mathcal I^{<}_{mn} = (-m)\,\mathbf 1_{m<0}\delta_{m+n,0}.
\end{equation}
Their difference is $m\delta_{m+n,0}$ for every $m\neq0$, and both terms vanish at $m=0$.

\subsection{Regulator and feature-width limits}

We keep the contour separation and retained mode window fixed in taking the following limits. At finite cutoff the current kernel is smooth and generally non-holomorphic.  The difference between two contour radii is controlled by its antiholomorphic derivative, as derived in appendix~\ref{app:contour-errors}.  Smooth spectral tapering permits the continuum limit on fixed, separated contours.  With $\ell_\Lambda=\log(\Lambda_{\rm UV}/\Lambda_{\rm IR})$, the lower and upper transition scales are $k_-=\Lambda_{\rm IR}e^{\tau\ell_\Lambda}$ and $k_+=\Lambda_{\rm IR}e^{(1-\tau)\ell_\Lambda}$. Both transitions must leave the physical window: $k_-\to0$ and $k_+\to\infty$.

The bound \eqref{eq:controlled-current-limit} separates the low-frequency error, proportional to $k_-^2$, from the high-frequency error, proportional to $k_+^{-1/2}$ at fixed contour separation.  If $V_{L,\Lambda,\tau}$ is the variance of one phase-averaged feature pairing, then with probability at least $1-\delta$ the normalized pairing error satisfies
\begin{equation}
\frac{\|\Omega_{N,\Lambda,\tau}-\Omega_0\|_F}{\|\Omega_0\|_F} \le\frac{C_{L,R}^{(R_0)}\varepsilon_{\rm reg}}{\|\Omega_0\|_F} +\frac1{\|\Omega_0\|_F} \sqrt{\frac{V_{L,\Lambda,\tau}}{N\delta}}.
\label{eq:pairing-limit-summary}
\end{equation}
The derivation is given in appendix~\ref{app:continuum-estimates}. At a fixed regulator the sampling term decreases with width. The bias vanishes along an admissible regulator path, giving convergence in the iterated limit. A joint limit holds when $V_{L,\Lambda,\tau}/N\to0$ along that path. 

\subsection{Fock representation and Virasoro action}
\label{sec:quantized}

The limits above give the canonical pairing on each fixed mode window. These limiting forms agree under restriction to smaller windows. On a common Fock space, take independent oscillators $u_n$, $n>0$, and set
\begin{equation}
[u_m,u_n^\dagger]=\delta_{mn},\qquad \widehat\alpha_n=\sqrt n\,u_n,\qquad \widehat\alpha_{-n}=\sqrt n\,u_n^\dagger,\qquad u_n|0\rangle=0. 
\label{eq:limiting-fock-oscillators}
\end{equation}
Finite products of creation modes generate a common dense domain. We specify radial conjugation on the unit circle in the normalized contour coordinate for the mode adjoint. The operators then obey
\begin{equation}
[\widehat\alpha_m,\widehat\alpha_n] = m\delta_{m+n,0}\mathbf1, \qquad \widehat\alpha_m^\dagger=\widehat\alpha_{-m}.
\label{eq:canonical-CCR}
\end{equation}

In the neutral sector, set $\widehat\alpha_0=0$. The modes assemble the chiral field
\begin{equation}
\begin{aligned}
\widehat J(z)&=\sum_{n\in\mathbb Z}\widehat\alpha_n z^{-n-1},\\ \langle0|\widehat J(z)\widehat J(w)|0\rangle &=\sum_{n>0}n z^{-n-1}w^{n-1} =\frac1{(z-w)^2},\qquad |z|>|w|.
\end{aligned}
\label{eq:fock-current-kernel}
\end{equation}
At fixed spectral cutoffs, the independent Fourier features have finite moments, and their large-width current correlations obey Wick's rule. Removing the regulator on separated insertions then gives
\begin{equation}
\begin{aligned}
\lim_{\Lambda}\lim_{N\to\infty} \E_\Theta\!\left[\prod_{i=1}^{2s}J_N(z_i)\right] &=\sum_{P\in\mathcal P_{2s}} \prod_{(i,j)\in P}\frac1{(z_i-z_j)^2}\\
&=\langle0|\mathcal R\!\left[\prod_{i=1}^{2s}\widehat J(z_i)\right]|0\rangle.
\end{aligned}
\label{eq:neural-fock-moment-identification}
\end{equation}
The set $\mathcal P_{2s}$ consists of pair partitions, and $\mathcal R$ denotes radial ordering. Odd correlations vanish. The outer limit follows \eqref{eq:admissible-regulator-path}; appendix~\ref{app:continuum-estimates} explains the moment limit. Thus the neural current determines the vacuum correlation functional of the Fock field. The positive-frequency splitting fixes its vacuum, and $\|\widehat\alpha_{-n}|0\rangle\|^2=n$ fixes the positive oscillator inner product. For local operator products we write $J=\widehat J$.

On a fixed charge sector, adjoin a commuting zero mode $\widehat\alpha_0=a\mathbf1$. For each $a\in\mathbb R$, the charged Fock module $\mathcal F_a$ is generated from
\begin{equation}
\widehat\alpha_0|a\rangle=a|a\rangle, \qquad \widehat\alpha_n|a\rangle=0 \quad(n>0).
\end{equation}
The Sugawara modes are
\begin{equation}
\widehat L_m = \frac12\sum_{p\in\mathbb Z} :\widehat\alpha_{m-p}\widehat\alpha_p:,
\label{eq:full-sugawara}
\end{equation}
and
\begin{equation}
\widehat L_0 = \frac12\widehat\alpha_0^2+ \sum_{p>0}\widehat\alpha_{-p}\widehat\alpha_p.
\end{equation}
Hence
\begin{equation}
\widehat L_0|a\rangle=\frac{a^2}{2}|a\rangle,
\label{eq:charged-weight}
\end{equation}
so $h_a=a^2/2$. Every charge sector has the same oscillator central term.

The local stress tensor of the canonically normalized current is
\begin{equation}
T^B(z)=\frac12:J(z)J(z):,\qquad T^B(z)T^B(w)\sim\frac{1/2}{(z-w)^4} +\frac{2T^B(w)}{(z-w)^2}+\frac{\partial T^B(w)}{z-w}.
\label{eq:bosonic-stress-ope}
\end{equation}
The double contraction fixes $c=1$, while the single contractions give the action on $T^B$ and the current.  Contour integration gives
\begin{align}
[\widehat L_m,\widehat\alpha_n]&=-n\widehat\alpha_{m+n},\nonumber\\
[\widehat L_m,\widehat L_n]&=(m-n)\widehat L_{m+n} +\frac{m(m^2-1)}{12}\delta_{m+n,0}\mathbf1.
\label{eq:virasoro-current-readout}
\end{align}

The central coefficient can also be read directly from the oscillator states. For $m>0$, normal ordering gives
\begin{equation}
\begin{aligned}
\widehat L_{-m}|0\rangle &=\frac12\sum_{p=1}^{m-1} \widehat\alpha_{-p}\widehat\alpha_{-(m-p)}|0\rangle,\\
\langle0|\widehat L_m\widehat L_{-m}|0\rangle &=\frac12\sum_{p=1}^{m-1}p(m-p) =\frac{m(m^2-1)}{12}.
\end{aligned}
\label{eq:central-term-from-neural-pairing}
\end{equation}
The current commutator in \eqref{eq:virasoro-current-readout} fixes the noncentral part of the stress-tensor commutator; the vacuum matrix element fixes its central term. A charge $a$ adds the highest weight $a^2/2$ through $\widehat\alpha_0$ while leaving this oscillator contribution unchanged.

At finite oscillator cutoff, the same action holds on each fixed charged level sector once the modes needed by that action are retained.  The explicit sufficient condition $K+m_-+n_-\le L$ and the treatment of edge terms are derived in appendix~\ref{app:finite-mode-support}. The current normalization fixed by the neural parameter measure thus determines both the central extension and the action of the Virasoro generators. The finite-window error bounds control the approach to this representation on the common finite-particle domain.

At finite width, compatibility with this adjoint and vacuum constrains the pairing. Order the nonzero modes as
\begin{equation}
(\alpha_1,\ldots,\alpha_L,\alpha_{-1},\ldots,\alpha_{-L})
\end{equation}
and write
\begin{equation}
C=\begin{pmatrix}0&I\\ I&0\end{pmatrix}, \qquad \alpha^\dagger=C\alpha, \qquad \Omega=\begin{pmatrix}A&G\\-G^T&B\end{pmatrix}.
\end{equation}
Compatibility of the central pairing with $\alpha_m^\dagger=\alpha_{-m}$ is the intrinsic relation
\begin{equation}
{C\overline\Omega C^T=-\Omega,}
\label{eq:intrinsic-star-condition}
\end{equation}
which implies $B=-\overline A$ and $G=G^\dagger$.

For a pairing satisfying \eqref{eq:intrinsic-star-condition}, an invertible canonicalization that preserves both the positive/negative mode polarization and this adjoint exists precisely when
\begin{equation}
{A=0,\qquad G=G^\dagger>0.}
\label{eq:physical-darboux-criterion}
\end{equation}
The canonical target has mixed block $D=\operatorname{diag}(1,\ldots,L)$. The change-of-basis argument and explicit canonicalizers are given in appendix~\ref{app:darboux}.

A generic empirical pairing has nonzero same-frequency blocks and a non-Hermitian mixed block. To define a finite-window oscillator readout we apply the explicit projection
\begin{equation}
H_N=\frac{G_N+G_N^\dagger}{2},\qquad \Omega_N^{\rm F}:=\mathcal P\Omega_N =\begin{pmatrix}0&H_N\\-H_N^T&0\end{pmatrix}.
\label{eq:fock-pairing-projection}
\end{equation}
This is an orthogonal projection in the real Frobenius inner product. Since $\mathcal P\Omega_0=\Omega_0$,
\begin{equation}
\|\Omega_N^{\rm F}-\Omega_0\|_F \le\|\Omega_N-\Omega_0\|_F.
\label{eq:fock-projection-contraction}
\end{equation}
We assign a Fock readout when $H_N>0$, using \eqref{eq:principal-physical-canonicalizer} with $G=H_N$. The projection specifies a change of the input pairing; its size and the raw defects are measured in section~\ref{sec:audits}. The kernel and Ward calculations use the original feature ensemble. 

The bound \eqref{eq:pairing-limit-summary} controls both the projected error and the probability of obtaining a positive $H_N$. Along its admissible limits, that probability tends to one and the finite readouts approach the limiting oscillators on each fixed level sector. Appendix~\ref{app:darboux} gives the positivity and matrix-element estimates.

Writing $\omega_N^{\rm F}$ for the form with matrix $\Omega_N^{\rm F}$, define the Heisenberg algebra by
\begin{equation}
[\widehat a(v),\widehat a(u)] = \omega_N^{\rm F}(v,u)\mathbf1.
\label{eq:matrix-CCR}
\end{equation}
Canonicalization gives the relations \eqref{eq:canonical-CCR}. For finitely many canonical pairs, every regular irreducible representation of the Weyl form of the CCR is unitarily equivalent by Stone--von Neumann uniqueness \cite{Folland:1989}.

\section{Conformal readouts and sectors}
\label{sec:readouts}

The current algebra and its Sugawara representation determine the charged fields and their Ward action. Fermionic and first-order architectures supply further conformal sectors, while reflected features relate the two chiral algebras at a boundary.

\subsection{Charged fields and Ward identities}
\label{sec:charged-sectors}
\label{sec:ward}

To connect the charged modules of section~\ref{sec:quantized}, introduce the coordinate $\widehat q$ conjugate to $\widehat\alpha_0$:
\begin{equation}
[\widehat q,\widehat\alpha_0]=i,
\qquad
[\widehat\alpha_0,\widehat\alpha_n]=0
\quad(n\neq0).
\end{equation}
One may work formally with a direct integral of charge sectors
$\int_{\mathbb R}^{\oplus}\mathcal F_a\,da$, where
$\widehat\alpha_0$ acts by multiplication by $a$ and
$\widehat q=i\partial_a$ on a common dense domain.  The operator $\widehat q$ moves between charge sectors. Within a
fixed module $\mathcal F_a$, $\widehat\alpha_0=a\mathbf1$.
The charged Fock states also determine the holomorphic weights of
the free-boson vertex operators.  For the real scalar,
the limiting non-chiral primaries are
\begin{equation}
V_a(z,\bar z)=:e^{ia\phi(z,\bar z)}:,
\qquad
h_a=\bar h_a=\frac{a^2}{2}.
\end{equation}
At finite feature width we define the corresponding composite observables
by covariance normal ordering with the same conditioned feature set,
\begin{equation}
V_{a,N}(w,\bar w)
=
\exp\left[
ia\phi_N(w,\bar w)
+\frac{a^2}{2}K_{\phi\phi}^{(N)}(w,w)
\right].
\label{eq:finite-network-vertex}
\end{equation}
Connected higher cumulants contribute at finite width. The
continuum primary also requires a convention for the infrared
constant. Choose a reference separation $\mu^{-1}$ and define
\begin{equation}
C_{\Lambda,\tau}^{(\mu)}:=K_{\Lambda,\tau}(\mu^{-1}),
\qquad
\mathcal V_{a,N}^{\Lambda,\tau}(w,\bar w)
=e^{-a^2C_{\Lambda,\tau}^{(\mu)}/2}V_{a,N}(w,\bar w).
\label{eq:infrared-renormalized-vertex}
\end{equation}
Equation~\eqref{eq:scalar-infrared-subtraction} gives
$K_{\Lambda,\tau}(r)-C_{\Lambda,\tau}^{(\mu)}
\to-\log(\mu^2r^2)$. At fixed regulator, the large-width limit gives
\begin{equation}
\lim_{N\to\infty}\E_\Theta\!\left[\prod_i V_{a_i,N}(z_i,\bar z_i)\right]
=\exp\!\left[-\sum_{i<j}a_i a_j K_{\Lambda,\tau}(z_i-z_j)\right].
\label{eq:gaussian-vertex-before-ir}
\end{equation}
For neutral charges, the constant part of this exponent is
$C_{\Lambda,\tau}^{(\mu)}\sum_i a_i^2/2$; the factor in
\eqref{eq:infrared-renormalized-vertex} removes it.
The vertex factors are independent of the current or stress
insertion and cancel from the Ward ratios used in the numerical
comparisons.

We define the continuum $V_a$ by the ordered large-width and
regulator limit of these renormalized fields. Measuring
coordinates in the reference units $\mu^{-1}$, their neutral
correlators are
\begin{equation}
\left\langle\prod_i V_{a_i}(z_i,\bar z_i)\right\rangle
=\prod_{i<j}|z_i-z_j|^{2a_i a_j},\qquad \sum_i a_i=0.
\label{eq:neutral-vertex-correlator}
\end{equation}
In particular
$\langle V_a(z,\bar z)V_{-a}(w,\bar w)\rangle
=|z-w|^{-2a^2}$, so the total scaling dimension is $\Delta_a=a^2$.
The local Ward action is
\begin{equation}
T(z)V_a(w,\bar w)\sim
\frac{a^2/2}{(z-w)^2}V_a(w,\bar w)
+\frac{\partial_w V_a(w,\bar w)}{z-w}.
\label{eq:vertex-stress-ope}
\end{equation}
The logarithmic scalar itself carries an infrared zero-mode ambiguity;
its currents and neutral vertex correlations have the stated conformal
transformation laws.  A compact target-space theory is specified by also choosing a charge
and winding lattice.

For neutral configurations, the current and stress-tensor Ward targets are
\begin{align}
R_J^{\rm CFT}(z)
&=
\sum_i\frac{a_i}{z-w_i},
\label{eq:current-ward}
\\
R_T^{\rm CFT}(z)
&=
\sum_i\left[
\frac{a_i^2/2}{(z-w_i)^2}
+
\frac{1}{z-w_i}
\sum_{j\ne i}\frac{a_i a_j}{w_i-w_j}
\right].
\label{eq:stress-ward}
\end{align}

In radial quantization, a charge-$a$ insertion at the origin and a
compensating charge at infinity give $\langle J(z)\rangle_a=a/z$
from \eqref{eq:current-ward}. This matches the zero mode
$\widehat\alpha_0=a\mathbf1$ of $\mathcal F_a$ and identifies the
charged states with the vertex construction.

The same stress tensor determines the cylinder vacuum energy.  For $z=e^u$,
the stress tensor transforms as
\begin{equation}
T_{\rm cyl}(u)=z^2T(z)-\frac{c}{24},\qquad
E_0=-\frac{\pi c}{6\ell}
\label{eq:cylinder-casimir}
\end{equation}
for equal left and right central charges $c$ on a cylinder of
circumference $\ell$ \cite{Blote:1986qm,Tong:2009np}.

\subsection{Fermionic and supersymmetric sectors}
\label{sec:dirac-readout}
\label{sec:susy}
\label{sec:ramond-readout}
\label{sec:fermionic-modes}

For the NS expansion $\psi(z)=\sum_r\beta_r z^{-r-1/2}$, the
Cauchy pole similarly fixes the ordered pairing.
For half-integer $r,s$, the two radial orderings define
\begin{align}
\mathcal J^{>}_{rs}
&=
\oint_{|z|=R_{\rm out}}\frac{dz}{2\pi i}\,z^{r-\frac12}
\oint_{|w|=R_{\rm in}}\frac{dw}{2\pi i}\,w^{s-\frac12}
K_{\psi\psi}(z,w),
\\
\mathcal J^{<}_{rs}
&=
\oint_{|z|=R_{\rm in}}\frac{dz}{2\pi i}\,z^{r-\frac12}
\oint_{|w|=R_{\rm out}}\frac{dw}{2\pi i}\,w^{s-\frac12}
K_{\psi\psi}(z,w).
\end{align}
With the fermionic exchange sign included, the ordering discontinuity is
\begin{equation}
\eta_{rs}=\mathcal J^{>}_{rs}-\mathcal J^{<}_{rs}
=\delta_{r+s,0},
\label{eq:fermionic-pairing}
\end{equation}
and canonical quantization gives
\begin{equation}
\{\widehat\beta_r,\widehat\beta_s\}
=\delta_{r+s,0}\mathbf1.
\label{eq:CAR}
\end{equation}

For the canonically completed Majorana field,
\begin{equation}
T^F=-\frac12:\psi\partial\psi:,
\qquad
\langle T^F(z)T^F(w)\rangle=\frac1{4(z-w)^4},
\qquad c_F=\frac12.
\label{eq:majorana-stress-kernel}
\end{equation}
The coefficient follows from the two graded Wick contractions of the
Cauchy kernel and its derivatives.  This supplies the free fermion sector
associated with the critical Ising theory \cite{Schultz:1964fv,DiFrancesco:1997nk}. The spin fields and a complete
modular-invariant Ising theory require additional sectors.

For the Dirac architecture \eqref{eq:dirac-architecture}, the neutral
current is
\begin{equation}
J_F=: \Psi^\dagger\Psi :=i:\psi_1\psi_2:,
\qquad J_F(z)J_F(w)\sim\frac1{(z-w)^2}.
\label{eq:dirac-current-ope}
\end{equation}
The minus sign from exchanging the two independent fermions cancels the
factor $i^2$.  The total Majorana stress tensor is equivalently
$T_1^F+T_2^F=\tfrac12:J_FJ_F:$, with $c=1$.  The LK scalar and the
Dirac construction therefore realize the same neutral chiral current
algebra and its Sugawara stress tensor.  This is the local current-sector
content of bosonization \cite{DiFrancesco:1997nk}.

Writing $J=i\partial H$ for the chiral boson, the charged fields
$:e^{\pm iH}:$ have $h=1/2$ and the same local singularities as the
Dirac fields \cite{Coleman:1974bu,Mandelstam:1975hb}. With $J_F=: \Psi^\dagger\Psi :$, $\Psi$ has charge $-1$
and corresponds locally to $:e^{-iH}:$.  The charged identification is completed by specifying a charge lattice
together with fermion cocycles and a spin structure.

Coupling an independent scalar to a Majorana field gives the
holomorphic free scalar multiplet.  A right-moving copy supplies the other chiral
sector of the free $\CN=(1,1)$ completion \cite{Friedan:1985ge}.

In the NS sector the local supercurrent is $G=:J\psi:$ and the total stress tensor is
$T=T^B+T^F$.  Wick contraction gives
\begin{equation}
G(z)G(w)\sim\frac1{(z-w)^3}+\frac{2T(w)}{z-w}.
\label{eq:free-supercurrent-ope}
\end{equation}
The two single contractions supply $2T^B$ and $2T^F$, while the
double contraction gives $2c/3=1$.
With
\begin{align}
\widehat L_n^B
&=\frac12\sum_p:\widehat\alpha_{n-p}\widehat\alpha_p:,
\\
\widehat L_n^F
&=\frac14\sum_r(2r-n):\widehat\beta_{n-r}\widehat\beta_r:,
\\
\widehat G_r
&=\sum_n\widehat\alpha_n\widehat\beta_{r-n},
\qquad
\widehat L_n=\widehat L_n^B+\widehat L_n^F,
\end{align}
the canonical readout obeys the NS algebra
\begin{align}
[\widehat L_m,\widehat G_r]
&=\left(\frac m2-r\right)\widehat G_{m+r},
\\
\{\widehat G_r,\widehat G_s\}
&=2\widehat L_{r+s}
+\frac{c}{3}\left(r^2-\frac14\right)
\delta_{r+s,0}\mathbf1,
\qquad c=\frac32.
\end{align}

In the Ramond sector the integer-mode algebra in \eqref{eq:ramond-mode-field} has
$b_0^2=1/2$ and $\{b_0,b_n\}=0$ for $n\ne0$.  An ordinary Grassmann
coefficient squares to zero, so this part of the Ramond readout must be
adjoined as a Clifford operator.  A parity-graded realization on
$\mathbb C^2\otimes\mathcal F_{\ne0}$ is
\begin{equation}
b_0=\frac{\sigma_1}{\sqrt2}\otimes P_{\ne0},\qquad
b_{n\ne0}=1\otimes\widetilde b_n,\qquad
(-1)^F=\sigma_3\otimes P_{\ne0},
\label{eq:ramond-clifford-readout}
\end{equation}
where $\widetilde b_n$ are canonical nonzero integer modes and $P_{\ne0}$
is their Fock-space parity.  The two
ground states form the smallest representation that retains fermion
parity \cite{Ginsparg:1988ui}.

Point splitting the Ramond kernel in \eqref{eq:ramond-ground-kernel}
against the NS short-distance subtraction gives
\begin{equation}
\langle T^F(z)\rangle_{\rm R}=\frac1{16z^2},\qquad
h_F^{\rm R}=\frac1{16},\qquad
h_{a}^{\rm R}=\frac{a^2}{2}+\frac1{16}.
\label{eq:ramond-ground-weight}
\end{equation}
The last weight includes the charged scalar. In a mode expression
normal ordered relative to a Ramond ground state, this shift is
\begin{equation}
\widehat L_n^{F,{\rm R}}
=\frac14\sum_{r\in\mathbb Z}(2r-n):
b_{n-r}b_r:_{\rm R}
+\frac1{16}\delta_{n,0}\mathbf1.
\label{eq:ramond-normal-ordering-shift}
\end{equation}
The ground-state expectation of the Ramond-normal-ordered quadratic
term vanishes; the additive term retains the plane stress-tensor
normalization. With this prescription, the integer-indexed
supercurrent obeys the same super-Virasoro relations, in particular
\begin{equation}
G_0^2=L_0-\frac{c}{24},\qquad c=\frac32.
\label{eq:ramond-supercharge}
\end{equation}
Thus the uncharged free multiplet has $h=c/24$.  

\subsection{First-order stress tensors and ghost sectors}
\label{sec:ghost-readouts}

We use algebraic first-order completions for the fields of
section~\ref{sec:first-order-architectures}. Choose weights $h_F=\lambda$ and $h_G=1-\lambda$.  The stress tensor is
\begin{equation}
T^{(\lambda)}=(1-\lambda):\partial F\,G:
-\lambda:F\,\partial G:.
\label{eq:first-order-stress}
\end{equation}
The pole conventions are
$F(z)G(w)\sim\kappa/(z-w)$ and $G(z)F(w)\sim1/(z-w)$, with
$\kappa=+1$ for anticommuting $bc$ fields and $\kappa=-1$ for commuting
$\beta\gamma$ fields.  These conventions give
\begin{align}
T^{(\lambda)}(z)F(w)&\sim\frac{\lambda F(w)}{(z-w)^2}
+\frac{\partial F(w)}{z-w},\nonumber\\
T^{(\lambda)}(z)G(w)&\sim\frac{(1-\lambda)G(w)}{(z-w)^2}
+\frac{\partial G(w)}{z-w}.
\label{eq:first-order-ward-weights}
\end{align}
The improvement parameter $\lambda$ assigns conformal weights to the
same local Cauchy pole. An oscillator vacuum and a zero-mode
prescription complete the representation.

Writing $A=1-\lambda$ and $B=-\lambda$, the double contractions give
$c/2=-\kappa(A^2+B^2+4AB)$.  Consequently
\begin{equation}
c_{bc}=1-3(2\lambda-1)^2,\qquad
c_{\beta\gamma}=3(2\lambda-1)^2-1.
\label{eq:ghost-central-charges}
\end{equation}
The reparametrization ghosts have $\lambda=2$ and $c_{bc}=-26$;
the commuting superghosts have $\lambda=3/2$ and
$c_{\beta\gamma}=11$ \cite{Friedan:1985ge,DiFrancesco:1997nk,Tong:2009np}.
The Gaussian variables in \eqref{eq:first-order-features} parametrize
ordered contractions. The ghost state space is supplied by the
first-order oscillator relations, while
\eqref{eq:physical-darboux-criterion} characterizes the positive scalar
Fock representation. The reparametrization and superghost readouts
use their standard nonunitary first-order representations.

To use these readouts in a string construction, one must specify the
ghost zero modes and BRST cohomology of the full state space. The central
charge balance gives a necessary condition:
\begin{equation}
c_{\rm tot}^{\rm bos}=D-26,\qquad
c_{\rm tot}^{\rm NSR}=\frac32D-26+11=\frac32D-15.
\label{eq:worldsheet-central-balance}
\end{equation}
These expressions vanish at $D=26$ and $D=10$, respectively.
Neural bosonic-string constructions and amplitudes are developed in
\cite{Frank:2026bui}.

\subsection{Conformal and superconformal boundary gluing}
\label{sec:boundary}

A reflecting boundary retains one conformal algebra by relating the two
chiral sectors \cite{Cardy:1984bb,Cardy:2004hm}.
The reflected construction also fixes the current gluing.  With
$J_{\sigma,N}=i\partial\Phi_{\sigma,N}$ and
$\bar J_{\sigma,N}=i\bar\partial\Phi_{\sigma,N}$, where the bar labels
right chirality, the boundary derivatives give
\begin{equation}
\left.J_{\sigma,N}\right|_{\mathbb R}
=\left.\sigma\bar J_{\sigma,N}\right|_{\mathbb R}.
\label{eq:boundary-current-gluing}
\end{equation}
This feature-level identity holds at every spectral cutoff.  In the limiting free-field representation, normal ordering
with the corresponding reflected contractions gives
\begin{equation}
\left.T^B\right|_{\mathbb R}
=\left.\bar T^B\right|_{\mathbb R},
\qquad T^B=\tfrac12:JJ:.
\label{eq:boundary-stress-gluing}
\end{equation}
The doubled stress tensor then defines one Virasoro algebra with $c=1$.
Equivalently, in the closed-channel boundary-state convention the local
condition is $(L_n-\bar L_{-n})|B\rangle=0$
\cite{Cardy:2004hm}. A consistent boundary state also requires a specified zero-mode
sector and annulus consistency.

The image term also fixes the dependence on distance to the boundary.
Subtracting scalar variances at two heights cancels the coincident bulk
term:
\begin{align}
V_\sigma(y;y_0)
&:=\mathbb E[\Phi_{\sigma,N}(iy)^2]
-\mathbb E[\Phi_{\sigma,N}(iy_0)^2]\nonumber\\
&=\sigma\bigl[K_{\Lambda,\tau}(2iy)
-K_{\Lambda,\tau}(2iy_0)\bigr]
\longrightarrow -2\sigma\log(y/y_0).
\label{eq:boundary-height-statistic}
\end{align}
The logarithmic slope is $-2\sigma$ for the normalization
$K=-\log|z-w|^2$.

The continuum fermion two-point functions form a matrix, with rows and columns
ordered as $(L,R)$:
\begin{equation}
S_\eta(z,w)=
\begin{pmatrix}
\displaystyle\frac1{z-w}&\displaystyle\frac{\eta}{z-\bar w}\\[6pt]
\displaystyle\frac{\eta}{\bar z-w}&
\displaystyle\frac1{\bar z-\bar w}
\end{pmatrix}.
\label{eq:boundary-fermion-matrix}
\end{equation}
In particular, the image term belongs to the mixed-chirality contraction.
The matrix satisfies $S_{ab}(z,w)=-S_{ba}(w,z)$, including the fermionic
exchange sign.  

For the fermion, take $z=x+i\varepsilon$ and $w=i\varepsilon$.  The mixed
correlator is
\begin{equation}
S_{LR}(z,w)=\frac{\eta}{x+2i\varepsilon},\qquad
|S_{LR}(z,w)|=\frac1{\sqrt{x^2+4\varepsilon^2}}.
\label{eq:boundary-fermion-geometry}
\end{equation}
The height controls the short-distance crossover. For
$|x|\gg2\varepsilon$, the magnitude approaches $1/|x|$.

For the independent free boson and fermion, set
\begin{equation}
T=T^B-\tfrac12:\psi_L\partial\psi_L:,
\qquad G=:J\psi_L:,
\label{eq:boundary-supercurrent}
\end{equation}
and define the right fields with the same phase conventions.  The two
reflection signs give
\begin{equation}
\left.T\right|_{\mathbb R}=\left.\bar T\right|_{\mathbb R},
\qquad
\left.G\right|_{\mathbb R}
=\left.\rho\bar G\right|_{\mathbb R},
\qquad \rho=\sigma\eta.
\label{eq:boundary-superconformal-gluing}
\end{equation}
Both $\rho=+1$ and $\rho=-1$ specify a preserved boundary
$\CN=1$ superconformal algebra with $c=3/2$
\cite{Gaberdiel:2008gk}.  In the convention $G=\bar G$, one chooses
$\eta=\sigma$; the opposite relative sign selects the other preserved
supercharge.  Here $\eta$ fixes the reflection law. Cylinder periodicity is fixed
independently by the spin sector; this construction uses the NS modes
of section~\ref{sec:susy}.

The mixed contractions of the limiting fields make this gluing explicit:
\begin{equation}
\langle J(z)\bar J(\bar w)\rangle=\frac{\sigma}{(z-\bar w)^2},
\qquad
\langle G(z)\bar G(\bar w)\rangle
=\frac{\sigma\eta}{(z-\bar w)^3}.
\label{eq:boundary-mixed-current-kernels}
\end{equation}
The supercurrent relation follows by Wick factorization of the
independent boson and fermion contractions.

\section{Numerical support}
\label{sec:audits}

We examine the chiral pairing and finite-width field correlations at
matched spectral regulators. Feature width, spectral cutoff, and
oscillator support are varied independently. Full numerical results,
methods and uncertainties, and finite-mode checks are given in
appendices~\ref{appnum-sec:audits},
\ref{appnum-app:numerical-methods}, and
\ref{appnum-app:formal-audit}, respectively.

\subsection{Current pairing and finite-window readouts}

\begin{figure}[t]
\centering
\includegraphics[width=0.94\linewidth]{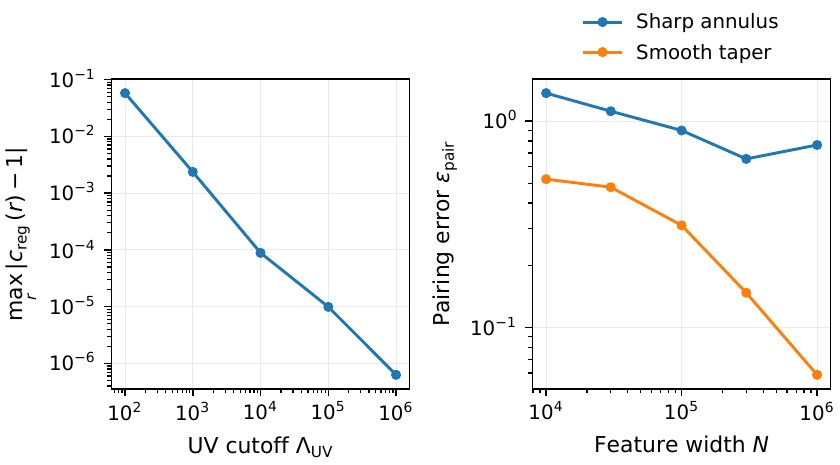}
\caption{Spectral and sampling effects. Left: maximum deviation of the analytic finite-regulator stress-tensor amplitude from its conformal value over the four insertion geometries, with $(\Lambda_{\rm IR},\Lambda_{\rm UV})=(b^{-1},b)$ and $\tau=0.15$. Right: empirical contour-pairing error at fixed regulator. Width reduces the sampling contribution while the spectral regulator determines the limiting bias.}
\label{fig:jhep-evidence-regulator-and-pairing}
\end{figure}

Smooth tapering is important for extracting the current modes because
differentiation amplifies the spectral endpoints. At the reference
regulator and contour geometry specified in
appendix~\ref{appnum-sec:regulator-architecture-checks}, the deterministic
relative pairing error is $0.00549$ for the raised-cosine window and
$0.786$ for the sharp annulus.
Figure~\ref{fig:jhep-evidence-regulator-and-pairing} compares the regulator
bias in the stress-tensor correlator with the empirical pairing error.

\begin{figure}[t]
\centering
\includegraphics[width=0.94\linewidth]{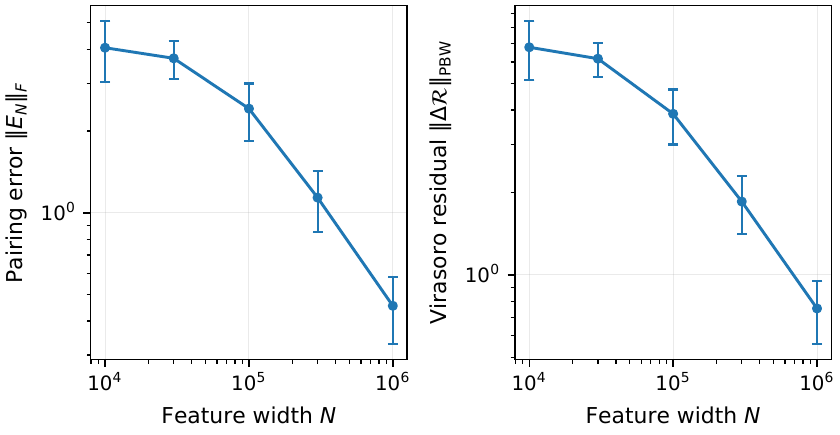}
\caption{Width dependence of the empirical pairing error and the Sugawara coefficient residual before canonicalization, with the canonical finite-mode remainder subtracted. The Log-Kernel architecture uses the tapered regulator. Points average four independent momentum samples.}
\label{fig:jhep-evidence-predarboux-width}
\end{figure}

Separating the empirical pairing from its regulated expectation gives
a sampling exponent $-0.4911\pm0.0157$, consistent with the width
dependence in \eqref{eq:pairing-limit-summary}. At fixed regulator,
the total error approaches the deterministic bias. The raw Sugawara
coefficient residual follows the pairing error as shown in
figure~\ref{fig:jhep-evidence-predarboux-width}, in agreement with the
finite-mode bound \eqref{eq:predarboux-stability}.

The saved empirical matrices also illustrate the finite Fock
qualification. All twenty have nonzero raw admissibility defects.
The projection \eqref{eq:fock-pairing-projection} gives a positive
mixed block in nineteen cases; one of the four samples at $N=10^4$
has a negative eigenvalue and remains a failed readout. At $N=10^6$,
the mean correction
$\|\Omega_N^{\rm F}-\Omega_N\|_F/\|\Omega_0\|_F$ is $0.00892$.
The positive readouts satisfy the canonical and adjoint identities
to numerical precision. Appendix~\ref{appnum-sec:audits} summarizes
these diagnostics; the repository retains the raw matrices, signed
eigenvalues, and projection corrections.

A phase-correlated control has a persistent translation defect even
though its contour pairing is close to the canonical target. This
shows why field symmetries are examined on the two-argument kernels
as well as on the extracted mode data. The architecture comparisons
and perturbation sweeps are given in
appendices~\ref{appnum-sec:regulator-architecture-checks}
and~\ref{appnum-app:formal-audit}.

\subsection{Composite and boundary correlations}
\label{sec:boundary-numerics}

The scalar four-point function follows the width dependence in
\eqref{eq:four-point-width-scaling}. At coincident insertions,
\begin{equation}
\begin{aligned}
\kappa_4(N)
&:=\E[\phi_N(x)^4]-3\E[\phi_N(x)^2]^2
=-\frac{3\sigma_\phi^4}{8N},\\
\gamma_2(N)
&:=\frac{\kappa_4(N)}{\E[\phi_N(x)^2]^2}=-\frac{3}{2N}.
\end{aligned}
\label{eq:four-point-equal-point-width}
\end{equation}
The saved tapered-window calculation gives a fitted exponent
$b_{\kappa_4}=-0.9951\pm0.0092$ over $8\le N\le128$,
consistent with the $1/N$ decay of the connected part.
Appendix~\ref{appnum-app:four-point-width} gives the contraction
formula and fitting convention.

The finite-network stress tensor uses conditional self-contraction
subtraction,
\begin{equation}
T_N(z)
=
\frac12\left[J_N(z)^2-K_{JJ}^{(N)}(z,z)\right].
\label{eq:finite-network-stress-tensor}
\end{equation}
Its directly sampled two-point function gives the matched-regulator
amplitude ratio $q_{\rm dir}=1.0083\pm0.0338$, using twelve independent
four-geometry block vectors. Precision varies with separation: the
largest-separation ratio is $1.920\pm0.562$. The Wick comparison,
connected-cumulant relation, geometry dependence, and correlated
uncertainties are given in
appendices~\ref{appnum-sec:composite-checks}
and~\ref{appnum-app:tt-statistics}.

Charged-field comparisons use the current and stress Ward targets
\eqref{eq:current-ward} and \eqref{eq:stress-ward}. For the stress
ratio, the relative RMS error over the prescribed probes is
\begin{equation}
\epsilon_T^{\rm relRMS}=
\left[\frac{\sum_{g\in\mathcal G}w_g
|\widehat R_{T,g}-R_{T,g}^{\rm CFT}|^2}
{\sum_{g\in\mathcal G}w_g|R_{T,g}^{\rm CFT}|^2}\right]^{1/2}.
\label{eq:ward-rel-rms}
\end{equation}
The weights are uniform, so targets near zero are included through
the pooled norm. The original covariance-based Gaussian comparison
at $N=2\times10^5$ has mean errors $0.60$--$0.97$ across the two-,
three-, and four-vertex configurations. Its improvement over an RBF
prior therefore occurs at appreciable absolute error.

Exact phase integration gives the finite-width cosine-ensemble
ratios at fixed momenta. In a separate neutral two-vertex, six-probe
benchmark at $N=2^{21}$ and
$(\Lambda_{\rm IR},\Lambda_{\rm UV},\tau)=(1/300,300,0.15)$,
the characteristic-function-weighted ensemble stress ratio has
relative RMS error $0.0239$, with a delete-one-block jackknife
standard error of $0.0099$ over eight independent momentum blocks.
The mean conditional error on those blocks is $0.1087\pm0.0103$.
The two estimates retain the distinction between a conditioned field
and an ensemble ratio. Appendix~\ref{appnum-app:ward-statistics}
derives the phase integrals and specifies the averaging and
uncertainty conventions.

For both fermionic constructions, the measured CAR and
super-Virasoro residual exponents before canonicalization are
compatible with $-1/2$. The boundary variance difference
\eqref{eq:boundary-height-statistic} gives a Neumann slope
$-1.991\pm0.008$ against the continuum value $-2$. The image-kernel
comparison also resolves the fermion crossover in
\eqref{eq:boundary-fermion-geometry}. The retained CAR support and
boundary calculations are described in
appendices~\ref{appnum-sec:boundary-numerics}
and~\ref{appnum-app:boundary-methods}.

\section{Discussion and outlook}
\label{sec:discussion}

The architecture and parameter measure determine a current contraction
whose radial-ordering discontinuity gives the Heisenberg pairing.
Its limiting current moments agree with the Fock vacuum
correlations, and the Sugawara stress tensor realizes the Virasoro
algebra with $c=1$. The central coefficient is fixed by the same
oscillator pairing. Smooth spectral regularization controls
the passage from the neural fields to the chiral kernel, while the
finite-mode estimates identify the oscillator support on which the
charged-sector action is exact. Together these give a construction
of conformal symmetry from neural data and quantitative control of
its approximation.

The Fock interpretation distinguishes the limiting representation
from a finite empirical pairing. The latter defines an auxiliary
oscillator readout after an explicit Hermitian mixed-block
projection, provided the resulting block is positive. The
projection error and positivity probability follow from the
pairing estimates, and fixed-level matrix elements converge on the
common Fock space. Field symmetries and Ward identities continue
to be tested on the original ensemble. The numerical comparisons
illustrate these statements, with the absolute Ward errors and
their regulator dependence recorded in the numerical appendices.

The sector extensions give the construction additional physical
content. Two Majorana fields reproduce the neutral bosonic current
algebra, and an independent scalar and Majorana field give the
free superconformal algebra. In the Ramond sector, the local
fermion pole agrees with the NS pole while the Clifford zero mode
changes the ground-state kernel and conformal weight. First-order
stress-tensor improvements vary weights and central charges at a
fixed Cauchy pole. Reflection acts on individual features and
imposes boundary conditions before averaging; its image kernels
give conformal and superconformal gluing. These boundary fields
provide a starting point for boundary critical phenomena and
quantum impurity problems with specified operator content
\cite{Cardy:1984bb,Affleck:1995ge}.

The canonical Hilbert space belongs to the limiting chiral
construction. At finite spectral cutoff the smooth, finite-variance
scalar ensemble falls within the reflection-positivity obstructions
discussed in \cite{FerkoMutchler:2026,Harvey:2026}. Continuum
reconstruction for the infrared-neutral or differentiated observables
considered here still requires the Osterwalder--Schrader conditions
\cite{Osterwalder:1973}. Higher
cumulants quantify departures from the Gaussian limit. In particular,
the connected scalar four-point function vanishes as $1/N$ at fixed
regulator, leaving the Wick correlations of the regulated free field. Interpreting
finite-width cumulants as local interactions would require an action
and control of locality.

The present estimates keep the contour separation and mode window
fixed. Extending them to increasing levels and to contours approaching
coincidence would give access to more singular observables.
The WZW and coset constructions of~\cite{CapuozzoRobinsonSuzzoni:2026}
extend this programme to non-Abelian currents and level--rank duality.

\section*{Data and code availability}

The code and computational evidence supporting this version are publicly
available in the repository~\cite{Robinson:2026VirasoroEvidence}.
The cited revision contains the numerical inputs and generation metadata,
with checksums and instructions for reproducing the reported statistics
and figures.

\acknowledgments

We thank Jim Halverson, Benjamin Suzzoni, and Pietro Capuozzo for comments on
earlier versions of the manuscript, and Pietro Capuozzo for advice on the
numerical experiments.  The smooth spectral-regulator framework and
ordered limiting prescription were developed jointly with Pietro Capuozzo
and Benjamin Suzzoni and also enter the WZW constructions
of~\cite{CapuozzoRobinsonSuzzoni:2026}. The raised-cosine current and pairing
estimates are established here. This work was supported by NWO Vidi grant
016.Vidi.189.182.

OpenAI ChatGPT and Codex, and Anthropic Claude Code, were used to assist with mathematical and numerical adversarial checks, code drafting
and refactoring, test generation, and release-provenance auditing.  All
mathematical and scientific assertions, computational outputs and citations
remain the responsibility of the author; the reproducibility tests and
release manifest provide the reported computational verification.  AI tools
are not treated as independent scientific sources or evidence.

\appendix
\section{Regulated kernels and contour estimates}
\label{app:regulator-details}
\subsection{Regulated kernels and feature expansions}
\label{app:kernel}
\label{app:mode-projection}
\label{app:angular-averaging}
\label{app:cauchy-sampling}

Phase averaging gives
\begin{equation}
K_{\Lambda,\tau}(r)
=
\frac{\sigma_\phi^2}{2\mathcal N_{\Lambda,\tau}}
\int_{\Lambda_{\rm IR}}^{\Lambda_{\rm UV}}
\frac{dk}{k}\,W_\tau(u(k))J_0(kr),
\qquad
\mathcal N_{\Lambda,\tau}
=\int_{\Lambda_{\rm IR}}^{\Lambda_{\rm UV}}
\frac{dk}{k}\,W_\tau(u(k)).
\label{eq:exact-regulated-scalar-kernel}
\end{equation}
For fixed positive separations $r,r_*$, the same normalization gives
\begin{equation}
\begin{aligned}
K_{\Lambda,\tau}(r)-K_{\Lambda,\tau}(r_*)
&=2\int_0^\infty \frac{dk}{k}\,W_\tau(u(k))
\bigl[J_0(kr)-J_0(kr_*)\bigr]
\\
&\longrightarrow -2\log(r/r_*).
\end{aligned}
\label{eq:scalar-infrared-subtraction}
\end{equation}
The Bessel difference divided by $k$ is integrable: it is $O(k)$
near zero and $O(k^{-3/2})$ at infinity. The expanding plateau makes
$W_\tau(u(k))\to1$ at each fixed $k>0$, so dominated convergence
removes the regulator. The limiting integral vanishes at $r=r_*$;
differentiation with Abel damping gives its derivative $-2/r$.
Choosing $r_*=\mu^{-1}$ gives the subtraction
used in \eqref{eq:infrared-renormalized-vertex}.

The current kernel is
\begin{equation}
K^{\Lambda,\tau}_{JJ}(z,w)
=-\partial_z\partial_w K_{\Lambda,\tau}(z-w).
\label{eq:exact-regulated-current-kernel}
\end{equation}
For the sharp window, with
$a=\Lambda_{\rm IR}r$ and $b=\Lambda_{\rm UV}r$, the derivative residuals
contain
\begin{equation}
J_0(b),
\qquad
2J_0(b)+bJ_1(b),
\qquad
\frac{b^2J_2(b)-a^2J_2(a)}{4u^2}.
\end{equation}
Their envelopes scale respectively as $b^{-1/2}$, $b^{1/2}$, and $b^{3/2}$.
Thus the scalar kernel and its first derivative can improve while the second
derivative and $\bar\partial K_{JJ}$ worsen.  The raised-cosine window removes
the discontinuous endpoint contribution.  Its $\bar\partial$ profile remains
oscillatory.  The quadrature resolves these oscillations, and radius-variation
errors are compared with a conservative Cauchy--Green envelope.

The same Fourier expansion determines the contour modes.  For
$k=|k|(\cos\theta_k,\sin\theta_k)$ and $z=re^{i\theta}$,
the Jacobi--Anger identity gives
\begin{equation}
    e^{ik\cdot x}
    =
    \sum_{n\in\mathbb Z}
    i^n J_n(|k|r)e^{in(\theta-\theta_k)}.
\end{equation}
After acting with the holomorphic derivative in
\eqref{eq:current-definition}, the contour projection
\eqref{eq:classical-mode-projection} becomes a finite feature map
\begin{equation}
    a_m^{(N)}(R;\Theta)
    =
    \sum_{j=1}^{N}F_{mj}(R;k_j,\theta_{k_j},\gamma_j),
\end{equation}
where each $F_{mj}$ is a Bessel-weighted Laurent coefficient.  The projection
map can be stored as a complex matrix $F$.  Its single-radius products
$FF^\dagger$ and $FF^T$ characterize projected covariances.  The chiral pairing
uses the two-radius functional \eqref{eq:operational-finite-pairing}, which
retains the radial-ordering information.  At finite cutoff the kernel is
non-holomorphic, so a single-radius covariance and the contour pairing carry
different information.

Angular averaging gives a complementary representation of these kernels.
At fixed radial momentum the angular part of the Fourier features can be
integrated exactly.  For a separation $re^{i\theta}$,
\begin{align}
\int_0^{2\pi}\frac{d\vartheta}{2\pi}
e^{-i\vartheta}\sin\bigl(kr\cos(\vartheta-\theta)\bigr)
&=e^{-i\theta}J_1(kr),\nonumber\\
\int_0^{2\pi}\frac{d\vartheta}{2\pi}
e^{-2i\vartheta}\cos\bigl(kr\cos(\vartheta-\theta)\bigr)
&=-e^{-2i\theta}J_2(kr).
\label{eq:angular-bessel-identities}
\end{align}
Thus the native Majorana and current kernels reduce respectively to
\begin{equation}
\begin{aligned}
K_{\psi\psi}^{\Lambda,\tau}(re^{i\theta})
&=e^{-i\theta}\int_{\Lambda_{\rm IR}}^{\Lambda_{\rm UV}}
W_\tau(u(k))J_1(kr)\,dk,\\
K_{JJ}^{\Lambda,\tau}(re^{i\theta})
&=\frac{e^{-2i\theta}}2\int_{\Lambda_{\rm IR}}^{\Lambda_{\rm UV}}
kW_\tau(u(k))J_2(kr)\,dk.
\end{aligned}
\label{eq:angular-averaged-kernels}
\end{equation}
Averaging over the angle removes that source of sampling noise. The
remaining radial integrals retain the spectral cutoffs; their quadrature
resolves the Bessel oscillations and supplies the reference kernel.

The discrete Cauchy features introduce a separate sampling law over
mode indices.  For a geometric proposal $q_n=(1-q)q^n$ on contour ratio
$r=|w/z|<1$, finite second moment requires
\begin{equation}
{q>r^2.}
\label{eq:cauchy-proposal-condition}
\end{equation}
Indeed
$\mathbb E_n|z^{-n-1}w^n/q_n|^2
=|z|^{-2}\sum_{n\ge0}r^{2n}/q_n$,
which is finite precisely when $q>r^2$ for the infinite-support proposal.
For a finite truncation, the proposal is renormalized on its support.
Contour extraction on a retained mode window inside that support is
unbiased for the truncated kernel. If $\widehat p_N(n)$ is the
empirical mode frequency and $q_n$ the actual proposal probability, then
\begin{equation}
\eta_{rs}^{(N)}=\delta_{r+s,0}
\frac{\widehat p_N(|r|-1/2)}{q_{|r|-1/2}}.
\label{eq:cauchy-empirical-pairing}
\end{equation}
Each selected amplitude has mean one and variance $(1-q_n)/(Nq_n)$.
A finite-window canonical CAR readout additionally requires every retained
mode to have been sampled: a missing mode gives a zero pairing amplitude and makes the retained
pairing degenerate.  For a fixed retained
index set $I$, the failure probability is at most
$\sum_{n\in I}(1-q_n)^N$.  The native boson-derived readout likewise requires
an admissible fermionic pairing in addition to the bosonic Fock
criterion.

\subsection{Contour deformations and radius dependence}
\label{app:contour-errors}
Suppose $K(z,w)$ is jointly meromorphic in a neighborhood of the product
of the swept annuli, holomorphic away from the diagonal, and has no other
singularity crossed in either contour deformation.  Then
\begin{equation}
\mathcal I^{>}_{mn}[K]-\mathcal I^{<}_{mn}[K]
=
\oint_{|w|=R}\frac{dw}{2\pi i}\,w^n
\oint_{|z-w|=\epsilon}\frac{dz}{2\pi i}\,z^m K(z,w).
\label{eq:residue-ordering-equivalence}
\end{equation}

Deform the outer $z$-contour through the inner one while keeping the
$w$-contour fixed.  By contour deformation, the difference between the two
nested contour integrals is the sum of the singularities crossed.  Under the
stated hypothesis this is precisely the small contour around $z=w$.

For a meromorphic chiral kernel, equation~\eqref{eq:residue-ordering-equivalence}
identifies the contour pairing with the usual local residue.  At finite
spectral cutoff the Euclidean current kernel is generally non-holomorphic,
and its dependence on both contour radii must be retained.

Let $\xi=z-w$, $u=\xi\bar\xi$, and write the regulated scalar kernel as
$K_\Lambda(z,w)=k_\Lambda(u)$. Differentiation gives
\begin{equation}
K_{JJ}^{\Lambda}(z,w)
=
(\bar z-\bar w)^2 k_\Lambda''(u),
\label{eq:radial-current-kernel-form}
\end{equation}
and
\begin{equation}
\partial_{\bar z}K_{JJ}^{\Lambda}(z,w)
=
(\bar z-\bar w)\left[2k_\Lambda''(u)+u k_\Lambda'''(u)\right].
\label{eq:dbar-current-kernel}
\end{equation}
On an interval of $u>0$, the current kernel is holomorphic in $z$ if and only
if
\begin{equation}
k_\Lambda(u)=-A\log u+B u+C
\end{equation}
there. The two holomorphic derivatives eliminate the affine terms
from $K_{JJ}^{\Lambda}$, and $A=1$ gives $(z-w)^{-2}$.

Equation~\eqref{eq:radial-current-kernel-form} follows from
$-\partial_z\partial_w k_\Lambda(u)$.  Applying $\partial_{\bar z}$ gives
\eqref{eq:dbar-current-kernel}.  Vanishing away from the diagonal is
$u k'''(u)+2k''(u)=0$, whose general solution is
$k(u)=-A\log u+B u+C$.

Thus the Bessel-regulated kernel is generically non-holomorphic at finite
cutoff.  Consequently, the extracted finite pairing depends genuinely on the
chosen inner and outer radii.  Two complementary bounds quantify this
dependence.

Let $K_0(z,w)=(z-w)^{-2}$ and
$\delta K=K_{JJ}^{(N,\Lambda)}-K_0$.  On the two contour tori
$T_{oi}=\{|z|=R_{\rm out},|w|=R_{\rm in}\}$ and
$T_{io}=\{|z|=R_{\rm in},|w|=R_{\rm out}\}$,
\begin{align}
\left|\omega_{mn}^{(N,\Lambda)}-m\delta_{m+n,0}\right|
&\le
R_{\rm out}^{m+1}R_{\rm in}^{n+1}
\|\delta K\|_{L^\infty(T_{oi})}
\nonumber\\
&\quad+
R_{\rm in}^{m+1}R_{\rm out}^{n+1}
\|\delta K\|_{L^\infty(T_{io})}.
\label{eq:finite-pairing-bound}
\end{align}

Subtract the ideal-kernel functionals and bound each normalized contour
integral by its contour length times the supremum norm.  The radii are
positive, so the expression applies to positive and negative integer modes.

For $R_{\rm out}' > R_{\rm out}>R_{\rm in}$ and a smooth finite-cutoff
kernel,
\begin{align}
&\left|
\mathcal I^>_{mn}(R_{\rm in},R_{\rm out}')
-
\mathcal I^>_{mn}(R_{\rm in},R_{\rm out})
\right|
\nonumber\\
&\qquad\le
\frac{R_{\rm in}^{n+1}}{\pi}
\iint_{A(R_{\rm out},R_{\rm out}')}
|z|^m
\sup_{|w|=R_{\rm in}}
\left|\partial_{\bar z}K_{JJ}^{(N,\Lambda)}(z,w)\right|
\,d^2z .
\label{eq:radius-variation-bound}
\end{align}
Analogous bounds hold when varying the inner contour and for
$\mathcal I^<$, with $z,m$ and $w,n$ exchanged.  The total variation of
$\omega_{mn}^{(N,\Lambda)}$ between two radius pairs is bounded by the sum of
the corresponding four terms.

Apply the Cauchy--Green formula to $z^mK_{JJ}^{(N,\Lambda)}(z,w)$ on the
annulus between the two outer contours and then take the normalized
$w$-contour integral.  Since $z^m$ is holomorphic on this annulus, only
$\partial_{\bar z}K_{JJ}^{(N,\Lambda)}$ contributes.

Equations~\eqref{eq:finite-pairing-bound} and
\eqref{eq:radius-variation-bound} relate contour dependence to the deviation from
the chiral kernel.  They provide complementary controls on the finite
regulator: proximity to the canonical pairing and stability as the radii are
varied. Numerical results for these bounds are described in
appendix~\ref{appnum-app:controls}.

\subsection{Smooth-regulator and sampling estimates}
\label{app:continuum-estimates}
\label{sec:pairing-limit}

We take the width limit before removing the raised-cosine regulator,
in the order used for the free-field constructions
of~\cite{CapuozzoRobinsonSuzzoni:2026}. Write
\begin{equation}
\ell_\Lambda=\log(\Lambda_{\rm UV}/\Lambda_{\rm IR}),\qquad
k_-:=\Lambda_{\rm IR}e^{\tau\ell_\Lambda},\qquad
k_+:=\Lambda_{\rm IR}e^{(1-\tau)\ell_\Lambda}.
\label{eq:effective-taper-scales}
\end{equation}
The raised-cosine window is identically one on $[k_-,k_+]$.  Its
regulator-removal limit must be taken along paths with
\begin{equation}
k_-\longrightarrow0,\qquad k_+\longrightarrow\infty,
\qquad 0<\tau<\tfrac12,
\label{eq:admissible-regulator-path}
\end{equation}
so that the plateau expands in both directions. For example,
$\Lambda_{\rm IR}=\mu e^{-T}$ and
$\Lambda_{\rm UV}=\mu e^{T}$ as $T\to\infty$ satisfy
\eqref{eq:admissible-regulator-path}.

Fix $0<R_{\rm in}<R_{\rm out}<\infty$ and a finite mode
window $\mathcal M_L^\times$.  Put
$d=R_{\rm out}-R_{\rm in}>0$ and let $W_\tau$ be the raised-cosine
window in \eqref{eq:raised-cosine-window}, with fixed
$0<\tau<1/2$.  If \eqref{eq:admissible-regulator-path} holds, then,
uniformly on both nested contour tori,
\begin{equation}
\|K_{JJ}^{\Lambda,\tau}-(z-w)^{-2}\|_\infty
\le \varepsilon_{\rm reg}(\Lambda,\tau;d)
:=M_\Lambda\left[\frac{k_-^2}{8}
+\frac{C_J}{2d^{5/2}\sqrt{k_+}}\right]
\longrightarrow0,
\label{eq:controlled-current-limit}
\end{equation}
provided $d k_+\ge1$, where
$M_\Lambda=2+\pi/(\tau\ell_\Lambda)$ and $C_J$ is any universal
constant such that $|J_0(x)|\le C_J x^{-1/2}$ for $x\ge1$.
For the pairing in $R_0$ units, set
$R_0=\sqrt{R_{\rm in}R_{\rm out}}$ and
$\rho=R_{\rm out}/R_{\rm in}>1$, and define
\begin{align}
C_{L,R}^{(R_0)}
&:=\left[\sum_{m,n\in\mathcal M_L^\times}
R_0^{-2(m+n)}\left(R_{\rm out}^{m+1}R_{\rm in}^{n+1}
+R_{\rm in}^{m+1}R_{\rm out}^{n+1}\right)^2\right]^{1/2}
\nonumber\\
&=R_0^2\left[\sum_{m,n\in\mathcal M_L^\times}
\left(\rho^{(m-n)/2}+\rho^{(n-m)/2}\right)^2\right]^{1/2}.
\label{eq:normalized-mode-bound-constant}
\end{align}
Then
\begin{equation}
{\|\Omega_{\infty,\Lambda,\tau}-\Omega_0\|_F
\le C_{L,R}^{(R_0)}\,\varepsilon_{\rm reg}(\Lambda,\tau;d).}
\label{eq:deterministic-pairing-convergence}
\end{equation}
The analogous unnormalized $z$-contour bound omits the factors
$R_0^{-(m+n)}$.  We use this normalization for the numerical pairings.

Equation~\eqref{eq:field-normalization} gives the exact spin-two kernel
for $\xi=z-w=re^{i\theta}$,
\begin{equation}
K_{JJ}^{\Lambda,\tau}(\xi)=
\frac{e^{-2i\theta}}{2}
\int_0^\infty kW_\tau(u(k))J_2(kr)\,dk.
\label{eq:exact-bessel-current}
\end{equation}
For brevity put $W(k)=W_\tau(u(k))$.  Twice using Bessel recurrence
relations and integrating by parts (the raised-cosine window has zero
boundary values and first derivatives) yields
\begin{equation}
\int_0^\infty kW(k)J_2(kr)\,dk
=\frac1{r^2}\int_0^\infty
\bigl[3W'(k)+kW''(k)\bigr]J_0(kr)\,dk.
\label{eq:bessel-ibp-identity}
\end{equation}
Derivatives may be understood piecewise at the window junctions:
$W$ and $W'$ are continuous and vanish at the outer endpoints.
In the logarithmic variable $u$ the integration-by-parts bracket obeys
\begin{equation}
\bigl[3W'(k)+kW''(k)\bigr]dk
=\left[2\,\partial_uW_\tau(u)
+\ell_\Lambda^{-1}\partial_u^2W_\tau(u)\right]du.
\end{equation}
On either taper ramp, the total variation bound is
$\int|3W'+kW''|\,dk\le2+\pi/(\tau\ell_\Lambda)$.
On the lower transition $[\Lambda_{\rm IR},k_-]$, the bracket has integral
$2$; on the upper transition $[k_+,\Lambda_{\rm UV}]$ it has integral
$-2$.  On the lower transition $|J_0(kr)-1|\le(kr)^2/4$, so its
contribution differs from $2/r^2$ by at most $M_\Lambda k_-^2/4$.
On the upper transition $kr\ge d k_+\ge1$, so
$|J_0(kr)|\le C_J/\sqrt{d k_+}$ and its contribution is at most
$M_\Lambda C_J/(d^{5/2}\sqrt{k_+})$.
Multiplying by $1/2$ as in \eqref{eq:exact-bessel-current} proves
\eqref{eq:controlled-current-limit}.  Applying
equation~\eqref{eq:finite-pairing-bound} entrywise, multiplying its
$(m,n)$ entry by the $R_0^{-(m+n)}$ factor in
\eqref{eq:r0-unit-pairing}, and summing squares proves
\eqref{eq:deterministic-pairing-convergence}.

For fixed $L,R_{\rm in},R_{\rm out},\Lambda,\tau$, let
$\Omega^{(1)}(k)$ be the contribution to the ordered pairing of one
phase-averaged momentum feature and put
\begin{equation}
V_{L,\Lambda,\tau}:=
\mathbb E_k\|\Omega^{(1)}(k)-\Omega_{\infty,\Lambda,\tau}\|_F^2.
\end{equation}
Then $V_{L,\Lambda,\tau}<\infty$ at finite spectral cutoffs and, for
$0<\delta<1$, with probability at least $1-\delta$,
\begin{equation}
{\|\Omega_{N,\Lambda,\tau}-\Omega_0\|_F
\le C_{L,R}^{(R_0)}\varepsilon_{\rm reg}(\Lambda,\tau;d)
+\sqrt{\frac{V_{L,\Lambda,\tau}}{N\delta}}.}
\label{eq:two-error-pairing-bound}
\end{equation}
The canonical pairing is reached in the iterated limit
$N\to\infty$ followed by any regulator path satisfying
\eqref{eq:admissible-regulator-path}.  A simultaneous triangular-array
limit holds in probability when $V_{L,\Lambda,\tau}/N\to0$ and the regulator
path satisfies \eqref{eq:admissible-regulator-path}.
For the selected, independently phase-averaged feature architecture,
$L_w=\mathcal Z_{\Lambda,\tau}/(2\pi)$ and the one-feature
current kernel obeys
$|K_{JJ}^{(1)}(z,w)|\le L_w\Lambda_{\rm UV}^2/2$.
Consequently the $R_0$-normalized single-feature pairing has
\begin{equation}
V_{L,\Lambda,\tau}
\le \frac14\left(C_{L,R}^{(R_0)}\right)^2
L_w^2\Lambda_{\rm UV}^4,
\label{eq:explicit-feature-variance-upper-bound}
\end{equation}
Thus $N/(L_w^2\Lambda_{\rm UV}^4)\to\infty$ is a sufficient
joint-width/cutoff growth condition at fixed mode and contour geometry.
It conservatively controls the growth of the single-feature variance
under regulator removal.

The phase-conditioned finite-feature pairing is a mean of $N$
independent identically distributed single-feature matrices.  Thus
\begin{equation}
\mathbb E\|\Omega_{N,\Lambda,\tau}-
\Omega_{\infty,\Lambda,\tau}\|_F^2
=\frac{V_{L,\Lambda,\tau}}{N}.
\end{equation}
Apply Markov's inequality to its squared Frobenius norm and use the
triangle inequality with equation~\eqref{eq:controlled-current-limit}.

Combined with equation~\eqref{eq:predarboux-stability}, this proves that,
on each fixed finite PBW window, the pairing-dependent raw Virasoro defect
tends to zero along the same admissible limits.  The observed
$N^{-1/2}$ fits in section~\ref{sec:audits} concern the sampling term in
\eqref{eq:two-error-pairing-bound} at fixed finite cutoffs.  The sufficient
joint-width condition separately controls the growth of its variance under
regulator removal.

The current moment identification uses the ordered width and regulator
limits. Write $J_N(z)=N^{-1/2}\sum_{j=1}^N j_j(z)$, where $j_j$
is the current of one feature with its regulated amplitude. At
fixed cutoffs these independent features have bounded moments.
Their connected cumulants satisfy
\begin{equation}
\kappa_r\!\left(J_N(z_1),\ldots,J_N(z_r)\right)
=N^{1-r/2}\,
\kappa_r\!\left(j_1(z_1),\ldots,j_1(z_r)\right).
\label{eq:current-cumulant-width-limit}
\end{equation}
The first cumulant vanishes, the second is the regulated current
kernel, and every cumulant of order greater than two vanishes as
$N\to\infty$. The moment--cumulant relation therefore gives Wick's
rule at each fixed regulator. On any finite set of separated
insertions, \eqref{eq:controlled-current-limit} then gives
\eqref{eq:neural-fock-moment-identification} term by term in the
finite Wick sum. Conditional covariance subtraction similarly
identifies the separated stress-tensor correlations with those
of $\tfrac12:\widehat J^2:$.

The joint-limit estimate \eqref{eq:two-error-pairing-bound} controls
the current pairing. For correlations of higher degree, a
simultaneous width and regulator limit also requires control of
the corresponding single-feature cumulants. The ordered limit
used in \eqref{eq:neural-fock-moment-identification} supplies that
correlation functional with no further growth assumption.
\section{Canonical representations and finite-mode estimates}
\label{app:representation-estimates}

\subsection{Finite-mode support}
\label{app:finite-mode-support}
For
\begin{equation}
v=
\widehat\alpha_{-r_1}\cdots\widehat\alpha_{-r_s}|a\rangle,
\qquad r_i>0,
\end{equation}
define $\ell(v)=\sum_i r_i$, and let $\mathcal F_a^{(\le K)}$ be the span of
states of level at most $K$.  The oscillator cutoff is denoted by $L$:
\begin{equation}
\widehat L_m^{(L)}
=
\frac12
\sum_{\substack{p\in\mathbb Z\\|p|\le L,\ |m-p|\le L}}
:\widehat\alpha_{m-p}\widehat\alpha_p:.
\label{eq:truncated-sugawara}
\end{equation}
Write $m_-=\max(0,-m)$.

If $v\in\mathcal F_a^{(\le K)}$ and
\begin{equation}
L\ge K+m_-,
\label{eq:single-mode-bound}
\end{equation}
then $\widehat L_m^{(L)}v=\widehat L_m v$.

Consider a normal-ordered summand
$:\widehat\alpha_{m-p}\widehat\alpha_p:$ acting nontrivially on $v$.  If both
indices are negative, the summand is pure creation, so $m<0$ and their total
created level is $m_-=|m|$.  If one index is zero, the other is $m$.  In every
remaining nonzero term there is a positive annihilation mode of index at most
$K$; for $m<0$ its partner has magnitude at most $K+m_-$.  Thus every
oscillator index needed on $\mathcal F_a^{(\le K)}$ lies inside the cutoff
under \eqref{eq:single-mode-bound}.

Let $m,n\in\mathbb Z$ and $K\ge0$.  If
\begin{equation}
{K+m_-+n_-\le L,}
\label{eq:filtered-bound}
\end{equation}
then, for every $v\in\mathcal F_a^{(\le K)}$,
\begin{equation}
[\widehat L_m^{(L)},\widehat L_n^{(L)}]v
=
(m-n)\widehat L_{m+n}^{(L)}v
+
\frac{1}{12}m(m^2-1)\delta_{m+n,0}\,v.
\label{eq:filtered-virasoro}
\end{equation}
Thus the truncated generators realize the $c=1$ Virasoro algebra exactly as
linear maps with domain $\mathcal F_a^{(\le K)}$.

The mode $\widehat L_n$ maps a level-$\ell$ vector to level $\ell-n$ when the
result is nonzero, and hence maps $\mathcal F_a^{(\le K)}$ into
$\mathcal F_a^{(\le K+n_-)}$.  The single-mode support bound and
\eqref{eq:filtered-bound} therefore replace each truncated mode in the two
ordered products by its full counterpart.  Moreover,
$(m+n)_-\le m_-+n_-$, so
$\widehat L_{m+n}^{(L)}v=\widehat L_{m+n}v$.  The standard free-boson relation
\cite{DiFrancesco:1997nk,GoddardOlive:1986,Sugawara:1968} then proves
\eqref{eq:filtered-virasoro}.

For fixed $K,m,n$, equation~\eqref{eq:filtered-virasoro} holds for every
sufficiently large $L$.  Along the natural inclusions of filtered Fock spaces,
$\widehat L_m^{(L)}$ converges strongly on each fixed filtered piece to
$\widehat L_m$.

\subsection{Pairing errors and coefficient residuals}
\label{app:pairing-residuals}
We express the Sugawara residual before canonicalization in a fixed
Poincar\'e--Birkhoff--Witt (PBW) basis of normally ordered monomials.  The
pairing error then controls its coefficient norm. Fix a finite mode set
containing the zero mode and choose a total PBW order refining the
positive/negative-mode splitting. This also fixes the order of modes
within each frequency block.  Let
\begin{equation}
L_m(\Omega)=\frac12\sum_{p\in P_m}:a_{m-p}a_p:
\end{equation}
be the finite-window quadratic expression, with central relations
$[a_r,a_s]=\Omega_{rs}\mathbf1$.  Define
\begin{equation}
\mathcal R_{m,n}(\Omega)
=
[L_m,L_n]_\Omega-(m-n)L_{m+n}
-\frac{m(m^2-1)}{12}\delta_{m+n,0}\mathbf1.
\label{eq:raw-virasoro-residual}
\end{equation}
Finite mode truncation can produce an $\Omega$-independent edge remainder.
The architecture-sensitive residual is therefore
\begin{equation}
\Delta\mathcal R_{m,n}(\Omega)
:=\mathcal R_{m,n}(\Omega)-\mathcal R_{m,n}(\Omega_0).
\label{eq:edge-subtracted-residual}
\end{equation}

Fix a mode cutoff $L$, filtered level $K$, and modes $m,n$.  In a fixed
normal-ordered PBW basis there are finite constants $A_{KmnL}$ and
$B_{KmnL}$ such that, for every antisymmetric perturbation $E$,
\begin{equation}
{
\|\Delta\mathcal R_{m,n}(\Omega_0+E)\|_{\rm PBW}
\le
A_{KmnL}\|E\|_F+B_{KmnL}\|E\|_F^2 .}
\label{eq:predarboux-stability}
\end{equation}
The coefficient norm in \eqref{eq:predarboux-stability} is taken
in this fixed algebraic basis. The filtered identity
\eqref{eq:filtered-virasoro} concerns the action of the canonical
remainder on $\mathcal F_a^{(\le K)}$. For example, at $L=2$,
$\mathcal R_{1,-1}(\Omega_0)=-3\widehat\alpha_{-2}
\widehat\alpha_2$: this nonzero PBW polynomial annihilates every
state of level at most one. The coefficient estimate therefore
controls the excess $\Delta\mathcal R$, while the support bound
controls the represented canonical action. At fixed $K,m,n,L$,
$\|E_N\|_F=O(N^{-1/2})$ gives
$\|\Delta\mathcal R_{m,n}(\Omega_N)\|_{\rm PBW}=O(N^{-1/2})$
before canonicalization.

For a Fock-admissible projected pairing, the common-space
realization below turns these finite oscillator polynomials
into maps between finite-level spaces. Their matrix elements
converge by \eqref{eq:fock-map-continuity}; the support condition
then supplies the limiting Virasoro action.

For central generator commutators,
\begin{equation}
[a_i a_j,a_k a_l]
=
a_i\Omega_{jk}a_l+a_i a_k\Omega_{jl}
+\Omega_{ik}a_l a_j+a_k\Omega_{il}a_j.
\end{equation}
Thus the commutator of two quadratic monomials has quadratic PBW coefficients
linear in $\Omega$.  Reordering to the fixed normal form introduces at most
one additional central commutator, so scalar coefficients are polynomial of
degree at most two.  Hence the finite coefficient map
$\Omega\mapsto\mathcal R_{m,n}(\Omega)$ is a polynomial of degree at most two.
Subtracting its value at $\Omega_0$ gives exactly
\begin{equation}
\Delta\mathcal R(\Omega_0+E)
=D\mathcal R|_{\Omega_0}[E]+Q(E,E),
\end{equation}
where $Q$ is bilinear.  In finite dimension both maps have finite operator
norms, giving \eqref{eq:predarboux-stability}.

Before canonicalization, write the bosonic and fermionic pairing errors as
$E_B$ and $E_F$.  The same PBW polynomial argument as in
equation~\eqref{eq:predarboux-stability} gives, after subtracting any
pairing-independent finite-mode edge remainder,
\begin{equation}
\begin{aligned}
\|\Delta\mathcal R_{\rm sVir}\|
&\le A_B\|E_B\|+A_F\|E_F\|\\
&\quad+B_{BB}\|E_B\|^2+B_{BF}\|E_B\|\|E_F\|+B_{FF}\|E_F\|^2.
\end{aligned}
\label{eq:supervir-stability}
\end{equation}
\subsection{Admissible finite-window readouts}
\label{app:darboux}

For the fixed polarization and mode adjoint of section~\ref{sec:quantized},
the target central form is
\begin{equation}
\Omega_{\rm can}
=
\begin{pmatrix}0&D\\-D&0\end{pmatrix},
\qquad D=\operatorname{diag}(1,\ldots,L).
\end{equation}
The conditions \eqref{eq:physical-darboux-criterion} characterize
canonicalization within this polarization.
When these conditions hold, a principal choice is
\begin{equation}
M_0=D^{1/2}G^{-1/2},
\qquad
S_0=\operatorname{diag}(M_0,\overline M_0),
\label{eq:principal-physical-canonicalizer}
\end{equation}
for which $S_0\Omega S_0^T=\Omega_{\rm can}$ and
$\overline S_0C=CS_0$.

A polarization-preserving map compatible with the involution has block form
$S=\operatorname{diag}(M,\overline M)$.  Its canonicalization equations are
\begin{equation}
MAM^T=0,\qquad MGM^\dagger=D.
\end{equation}
Invertibility requires $A=0$ and implies that $G$ is Hermitian positive
definite.  Conversely, if $A=0$ and $G>0$, the choice
\eqref{eq:principal-physical-canonicalizer} gives
$M_0GM_0^\dagger=D$ and automatically preserves the involution.

For an antisymmetric pairing
$\Omega=\bigl(\begin{smallmatrix}A&G\\-G^T&B\end{smallmatrix}\bigr)$,
the nearest matrix with vanishing same-frequency blocks and a Hermitian
mixed block is $\mathcal P\Omega$ in
\eqref{eq:fock-pairing-projection}. Indeed, the Hermitian and
anti-Hermitian parts of $G$ are orthogonal in the real Frobenius
inner product. With $E=\Omega-\Omega_0$, this gives
\begin{equation}
\|E\|_F^2
=\|\mathcal P\Omega-\Omega_0\|_F^2
+\|\Omega-\mathcal P\Omega\|_F^2.
\label{eq:fock-projection-pythagoras}
\end{equation}
The projection removes $A,B$ and the anti-Hermitian part of $G$.
It preserves the canonical target and inherits its bias and sampling
bounds.

Put
\begin{equation}
h_N=D^{-1/2}H_ND^{-1/2},\qquad
\eta_N=\|h_N-I\|_{\rm op}
\le\frac{\|\Omega_N-\Omega_0\|_F}{\sqrt2}.
\label{eq:fock-positivity-certificate}
\end{equation}
The last inequality uses $\lambda_{\min}(D)=1$ and the two
mixed blocks in the Frobenius norm. If $\eta_N<1$, then
$H_N\ge(1-\eta_N)D>0$. An explicit canonicalizer with a useful
stability estimate is
\begin{equation}
\widetilde M_N=D^{1/2}h_N^{-1/2}D^{-1/2},\qquad
\widetilde S_N=\operatorname{diag}
(\widetilde M_N,\overline{\widetilde M_N}).
\label{eq:normalized-fock-canonicalizer}
\end{equation}
It obeys
$\widetilde S_N\Omega_N^{\rm F}\widetilde S_N^T=\Omega_0$
and preserves the mode adjoint. Spectral calculus gives
\begin{align}
\|D^{-1/2}\widetilde M_ND^{1/2}-I\|_{\rm op}
&\le(1-\eta_N)^{-1/2}-1,\nonumber\\
\|D^{-1/2}\widetilde M_N^{-1}D^{1/2}-I\|_{\rm op}
&\le1-\sqrt{1-\eta_N}.
\label{eq:fock-map-continuity}
\end{align}
Thus the finite readout may be realized on the common limiting Fock
space by $\widehat a_N^{\rm F}
=\widetilde S_N^{-1}\widehat\alpha$. At fixed $L$, oscillator
polynomials of fixed degree have continuous matrix elements between
fixed finite-level vectors as $\eta_N\to0$. Choosing
$L\ge K+m_-+n_-$ then gives the charged-sector Virasoro action in
\eqref{eq:filtered-virasoro}.

To quantify the probability of a positive readout, write
$B_{\Lambda}=C_{L,R}^{(R_0)}\varepsilon_{\rm reg}$. If
$B_{\Lambda}<\sqrt2$, the variance estimate in
appendix~\ref{app:continuum-estimates} implies
\begin{equation}
\mathbb P\!\left[\lambda_{\min}(H_N)\le0\right]
\le
\frac{V_{L,\Lambda,\tau}}
{N(\sqrt2-B_{\Lambda})^2}.
\label{eq:fock-admissibility-probability}
\end{equation}
A nonpositive $H_N$ requires
$\|\Omega_N-\Omega_0\|_F\ge\sqrt2$. The triangle inequality
therefore bounds this event by
$\|\Omega_N-\Omega_{\infty,\Lambda,\tau}\|_F
\ge\sqrt2-B_{\Lambda}$, and Markov's inequality gives
\eqref{eq:fock-admissibility-probability}. On the regulator paths
and width scalings already specified, the right-hand side tends to
zero. Samples with a nonpositive $H_N$ are recorded as failures of
this finite readout; their eigenvalues are retained.
\section{Numerical results}
\label{appnum-sec:audits}

Each numerical observable is compared with its analytic reference at
the same spectral regulator. The feature width $N$, oscillator cutoff
$L$, and number $M$ of sampled networks are varied independently.
Section~\ref{appnum-app:numerical-methods} gives the numerical procedures and
uncertainty calculations.
\subsection{Regulator and architecture dependence}
\label{appnum-sec:regulator-architecture-checks}

The scalar covariance has a logarithmic scaling window for both the sharp
and tapered regulators.  Their current kernels behave differently because
the derivatives amplify spectral endpoints.  For a sharp annulus, the
current-kernel endpoint has an oscillatory envelope proportional to
$(\Lambda_{\rm UV}r)^{1/2}$, and the non-holomorphy measure is still more
sensitive to the UV boundary.  The raised-cosine taper suppresses this
contribution while preserving the interior $|k|^{-2}$ law.

For the reference comparison we use
$(\Lambda_{\rm IR},\Lambda_{\rm UV})=(10^{-2},10^2)$,
$R_0=1$, $\rho=2$, and retained mode cutoff $L=4$.
The deterministic contour-pairing errors are
\begin{equation}
\epsilon_{\rm pair}^{\rm sharp}=0.786,
\qquad
\epsilon_{\rm pair}^{\rm taper}=0.00549.
\end{equation}
The smooth regulator thus substantially reduces the departure from the
chiral pairing at these finite cutoffs. Its continuum behavior is governed
by equation~\eqref{eq:controlled-current-limit}, along paths satisfying
\eqref{eq:admissible-regulator-path}.  Figure~\ref{fig:jhep-evidence-regulator-and-pairing}
shows the regulator dependence alongside the empirical pairing error.

To separate sampling fluctuations from regulator bias, we compare the
empirical pairing with both its regulated expectation and the canonical
target.  Let $L$ denote the oscillator/mode cutoff.  The spectral cutoffs are
$\Lambda_{\rm spec}=(\Lambda_{\rm IR},\Lambda_{\rm UV})$.
On
\begin{equation}
\mathcal M_L^\times=\{-L,\ldots,-1,1,\ldots,L\},
\end{equation}
write the matrix of the contour pairing as
\begin{equation}
\Omega_N=\Omega_0+E_N,
\qquad
(\Omega_0)_{mn}=m\delta_{m+n,0}.
\end{equation}
The principal pairing errors are
\begin{align}
\epsilon_{\rm pair}
&=
\frac{\|\Omega_N-\Omega_0\|_F}{\|\Omega_0\|_F},
\\
\epsilon_{\rm samp}
&=
\frac{\|\Omega_N-\Omega_\infty\|_F}{\|\Omega_0\|_F},
\\
\epsilon_{\rm leak}
&=
\frac{\bigl(\sum_{m+n\neq0}|(\Omega_N)_{mn}|^2\bigr)^{1/2}}
{\bigl(\sum_{m+n=0}|(\Omega_N)_{mn}|^2\bigr)^{1/2}}.
\label{appnum-eq:pairing-errors}
\end{align}
For a rotationally invariant analytic kernel the pairing is exactly
antidiagonal, so analytic leakage vanishes structurally.  Leakage therefore measures angular contamination in a finite feature sample.

Sampling independently across feature sets gives
\begin{equation}
b_{\rm samp}=-0.4911\pm0.0157,
\qquad
R^2=0.953,
\qquad
p(b=-1/2)=0.57,
\label{appnum-eq:sampling-exponent}
\end{equation}
for the tapered regulator.  Over the measured width range, the total pairing
error is compatible with the same power law.  Its infinite-width limit at
fixed regulator is the deterministic pairing error, equal to $0.00549$ at
the reference geometry.  The sharp-window sampling term also follows the
Monte Carlo law, while its total error approaches a much larger regulator
floor.  

The bound transfers an independently established pairing estimate to
the Sugawara residual.  We examine its dependence on perturbation size in a
five-decade sweep and compare it with the finite-width calculations.
Along the reference perturbation direction the ratio
\begin{equation}
\frac{\|\Delta\mathcal R\|}
{\|E\|_F+\|E\|_F^2}
\end{equation}
lies between $1.6502$ and $1.8151$; across twelve perturbation directions the
corresponding band is $1.5583$--$2.1106$.  The perturbation sweep holds the mode window fixed.  The width
comparison instead samples the pairing errors induced by the architecture.

We vary the radial spectrum, angular distribution and phase correlations
of the features. Mode mixing and sign changes act directly on the
pairing. Field invariance is tested on the two-argument kernel, and the
projected modes are tested with the Fock diagnostics below.

We quantify departures from these conditions directly on $\Omega$:
\begin{align}
\delta_\star
&=
\frac{\|C\overline\Omega C^T+\Omega\|_F}{\|\Omega\|_F},
\\
\delta_{\rm iso}
&=
\frac{\|A\|_F}{\|\Omega\|_F},
\\
\delta_H
&=
\frac{\|G-G^\dagger\|_F}{\|G\|_F},
\label{appnum-eq:intrinsic-admissibility-defects}
\end{align}
together with the signed minimum eigenvalue of
$D^{-1/2}(G+G^\dagger)D^{-1/2}/2$ and the canonical-amplitude error
\begin{equation}
\delta_G :=\frac{\|D^{-1/2}GD^{-1/2}-I\|_F}{\sqrt L}.
\label{appnum-eq:target-crossblock-deviation}
\end{equation}
Here $\delta_{\rm iso}$ measures the positive-frequency block, while
$\delta_G$ compares the cross-block normalization with the target.
All twenty tapered Log-Kernel matrices in the saved width grid have
nonzero admissibility defects. At $N=10^6$, the mean values are
$\delta_{\rm iso}=0.00603$ and $\delta_H=0.00760$.

Applying \eqref{eq:fock-pairing-projection} gives a positive mixed
block in three of the four samples at $N=10^4$ and all four samples
at each larger width. The remaining low-width sample has a negative
eigenvalue and is recorded as a failed readout. At $N=10^6$, the mean
raw and projected pairing errors are $0.05877$ and $0.05800$,
respectively; the mean projection correction is
$\|\Omega_N^{\rm F}-\Omega_N\|_F/\|\Omega_0\|_F=0.00892$.
The nineteen positive readouts satisfy the canonical and adjoint
identities to numerical precision. This calculation uses the same
saved matrices as the raw diagnostics.

Translation invariance is measured on the scalar and current kernels as
functions of both insertion points.  For base pairs $(x_g,y_g)$ and common
translations $a$, define
\begin{equation}
\delta_{\rm TI}
=
\left[
\frac{\sum_{g,a}|K(x_g+a,y_g+a)-K(x_g,y_g)|^2}
{\sum_{g,a}|K(x_g,y_g)|^2}
\right]^{1/2}.
\label{appnum-eq:translation-defect}
\end{equation}
For the logarithmic architecture and the six independent-phase controls,
the scalar defect is at most $1.728\times10^{-16}$ and the current defect
at most $4.840\times10^{-15}$ over the measured widths. Correlated phases
instead give scalar defects close to $0.23$ across a $100\times$ width
range, although their contour pairing approximates the target. The resulting kernel depends on the center-of-mass coordinate as
well as $z-w$.

Rotational invariance is recovered statistically as the empirical angular
measure approaches its isotropic prior.  The rotational defect of the
logarithmic architecture has fitted exponent $-0.496$.  For an anisotropic
angular prior and angular-mode scrambling, the exponents are $-0.005$ and
$-0.018$, respectively, indicating persistent angular dependence.

The pairing and Fock conditions distinguish further changes.  For the RBF
prior, the normalized cross-block error $\delta_G$ lies between $1.373$ and
$1.379$ across two decades of width.  This is a mismatch with the canonical
amplitude; Fock compatibility is tested separately by the conditions on $A$
and $G$.  The local-exponent, sign-breaking, and non-symplectic controls have
negative normalized minimum eigenvalues, with representative values
$-0.644$, $-1.158$, and $-11.44$.  Their pairings therefore fail the positivity
condition for the chosen Fock polarization.

The pairing and symmetry fits are consistent with the expected
$N^{-1/2}$ sampling law for the logarithmic architecture. An equal-weight
average of eight fitted exponents gives $-0.4981$, with a shared-seed
cluster-bootstrap 95\% interval $[-0.5331,-0.4663]$. The resampling accounts
for dependence between observables evaluated on the same feature sets.
The fitting and pooling conventions are given in
section~\ref{appnum-app:width-fits}.

\begin{figure}[t]
\centering
\includegraphics[width=0.94\linewidth]{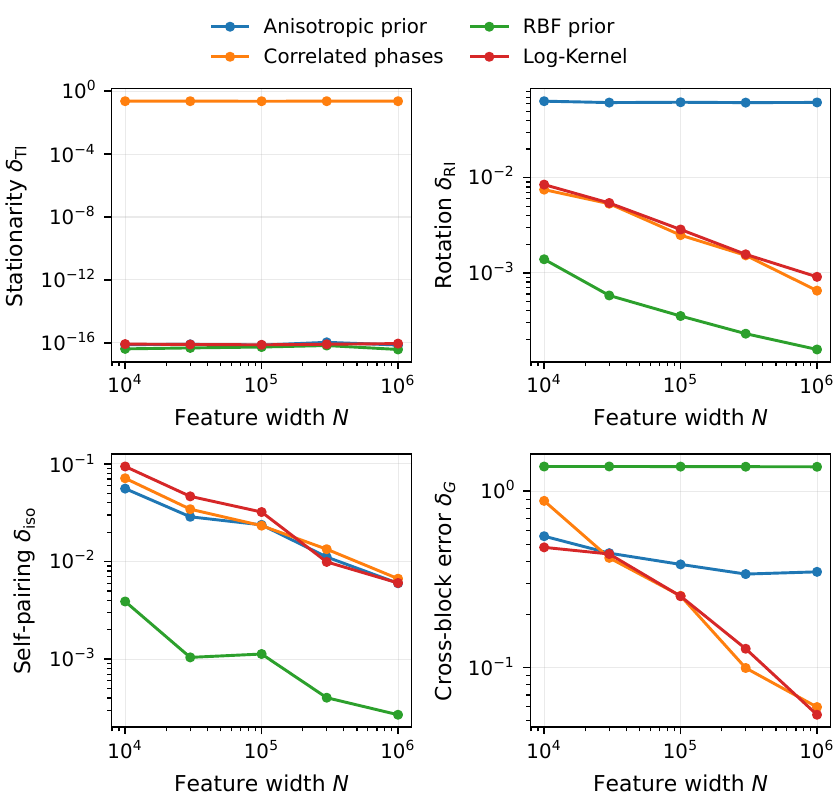}
\caption{Dependence on architecture. Translation invariance is evaluated on the two-argument kernel, while rotational errors measure the sampled angular distribution. The lower panels distinguish the positive-frequency-block and normalized cross-block errors.}
\label{appnum-fig:jhep-evidence-architecture-symmetries}
\end{figure}

\subsection{Composite correlations and Ward identities}
\label{appnum-sec:composite-checks}

\begin{figure}[t]
\centering
\includegraphics[width=0.94\linewidth]{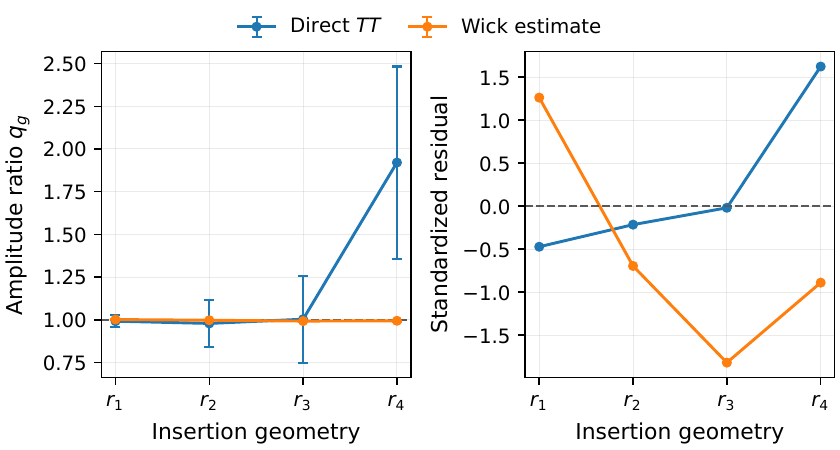}
\caption{Geometry dependence of the direct and Wick stress-tensor ratios and their fit-centered residuals. Uncertainty is evaluated over twelve independent blocks with correlated geometries. The largest direct separation is poorly determined; covariance tests are specified in section~\ref{appnum-app:tt-statistics}.}
\label{appnum-fig:jhep-evidence-tt-geometry}
\end{figure}

For the $c=1$ readout, we estimate the stress-tensor correlation using
the finite-network field with conditional self-contraction subtraction:
\begin{equation}
T_N(z)
=
\frac12\left[J_N(z)^2-K_{JJ}^{(N)}(z,z)\right].
\label{appnum-eq:finite-network-stress-tensor}
\end{equation}
We use both a Wick-factorized estimate, built from the current two-point
function, and a direct four-field estimate,
\begin{align}
\widehat G_{TT}^{\rm Wick}(z,w)
&=\frac12\widehat K_{JJ}(z,w)^2,
\\
\widehat G_{TT}^{\rm dir}(z,w)
&=\frac1M\sum_{a=1}^{M}T_{N,a}(z)T_{N,a}(w).
\end{align}
For the isotropic ensemble, the population expectations differ by the
unconditional connected current cumulant,
\begin{equation}
G_{TT}^{\rm dir}-G_{TT}^{\rm Wick}
=\frac14\kappa_{JJJJ}(z,z,w,w).
\label{appnum-eq:direct-wick-cumulant}
\end{equation}
The cancellation of the conditional self-contraction covariance in
this relation is derived in section~\ref{appnum-app:finite-width}.
Both are compared with the analytic finite-regulator reference at the same
insertion geometry. We define the normalized amplitude ratios by
\begin{equation}
q_g
=
\frac{c_{{\rm estimator},g}}
{c_{{\rm analytic,reg},g}}.
\label{appnum-eq:tt-regulated-ratio}
\end{equation}
The generalized least-squares summaries are
\begin{equation}
{
q_{\rm dir}=1.00830\pm0.03377,
\qquad
q_{\rm Wick}=0.99941\pm0.00133,}
\label{appnum-eq:tt-hardening-result}
\end{equation}
The uncertainties are plug-in standard errors from twelve independent
four-geometry block vectors.  Joint Hotelling tests of the four mean ratios
against one give $p=0.246$ for the direct estimator and $p=0.319$ for the Wick
estimator, under an approximately Gaussian block model.  The direct ratio at the largest separation is $q=1.920\pm0.562$,
with substantially greater uncertainty than at smaller separations.
Figure~\ref{appnum-fig:jhep-evidence-tt-geometry} displays the geometry dependence.
The covariance calculation, geometry table, and common-ratio contrast test
are given in section~\ref{appnum-app:tt-statistics}.

The regulated stress-tensor correlator approaches its conformal limit
along the spectral paths in \eqref{eq:controlled-current-limit}.

The field cumulant gives another way to follow this finite-width contribution.  For the cosine ensemble the connected equal-point fourth cumulant is
$\kappa_4=-3\sigma_\phi^4/(8N)$.  A fit with a free width exponent gives
\begin{equation}
b_{\kappa_4}=-0.9951\pm0.0092,
\qquad R^2=0.9997,
\end{equation}
consistent with the $1/N$ decay of the departure from Gaussianity.  This
scalar statistic measures Gaussianity. The separated derivative
cumulant in \eqref{appnum-eq:direct-wick-cumulant} also depends on the
momentum factors carried by the currents.

The charged-sector comparison uses the current and stress-tensor Ward
targets in \eqref{eq:current-ward} and \eqref{eq:stress-ward},
evaluated analytically at the probe points.  For a block
of $M$ independent network realizations, write
$X_{0,a}=\prod_iV_{a_i,N}^{(a)}(w_i)$ and form
\begin{equation}
\widehat R_J^{(b)}
=
\frac{\overline{J_NX_0}^{(b)}}{\overline X_0^{(b)}},
\qquad
\widehat R_T^{(b)}
=
\frac{\overline{T_NX_0}^{(b)}}{\overline X_0^{(b)}}.
\label{appnum-eq:sampled-ward-ratios}
\end{equation}
The numerator and denominator are evaluated on the same networks, retaining
their covariance. We measure the stress-Ward deviation over a fixed probe
set $\mathcal G$ using uniform weights $w_g=1$ in the relative RMS error
\begin{equation}
\epsilon_T^{\rm relRMS}=
\left[\frac{\sum_{g\in\mathcal G}w_g
|\widehat R_{T,g}-R_{T,g}^{\rm CFT}|^2}
{\sum_{g\in\mathcal G}w_g|R_{T,g}^{\rm CFT}|^2}\right]^{1/2}.
\label{appnum-eq:ward-rel-rms}
\end{equation}
Pooling the squared errors over the probe set avoids division by individual
Ward targets near zero.

\begin{figure}[t]
\centering
\includegraphics[width=0.94\linewidth]{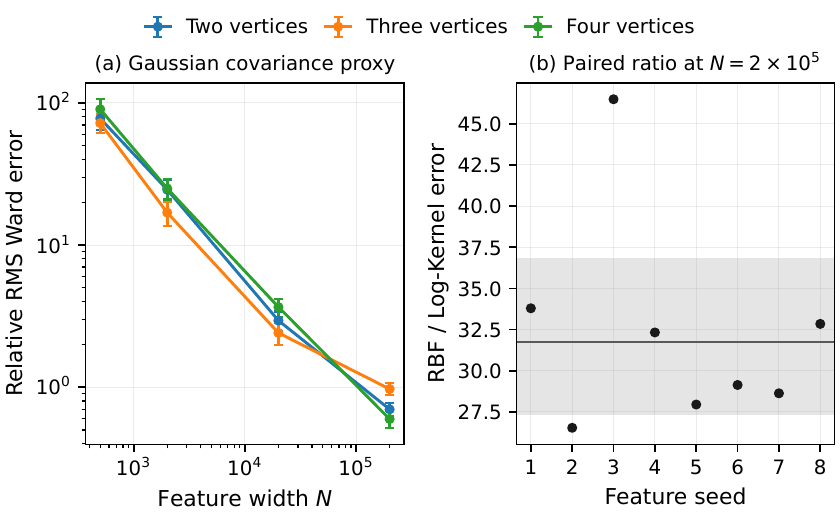}
\caption{Absolute and relative accuracy of covariance-based Gaussian Ward
functions. Left: mean relative RMS stress-Ward errors for the two-, three-
and four-vertex configurations of the Log-Kernel calculation, with standard
errors over eight feature seeds. Right: paired RBF-to-Log-Kernel error
ratios at $N=2\times10^5$. Each point averages log ratios over the three
configurations for one independent seed. The solid line is their
geometric mean; shading gives its Student-$t_7$ interval. These functions
are evaluated from the empirical covariance using
\eqref{appnum-eq:gaussian-ward-proxy}.}
\label{appnum-fig:jhep-evidence-ward-independent-blocks}
\end{figure}

The architecture comparison evaluates the Gaussian Ward
functions of the empirical covariance,
\begin{equation}
R_{J,N}^{\rm G}(z)=-\sum_i a_i\partial_z K_{\phi\phi}^{(N)}(z,w_i),
\qquad
R_{T,N}^{\rm G}(z)=\frac12\bigl[R_{J,N}^{\rm G}(z)\bigr]^2.
\label{appnum-eq:gaussian-ward-proxy}
\end{equation}
At finite cosine width these give a covariance-based approximation
to the phase-integrated ratios. For the tapered regulator at
$N=2\times10^5$, the mean relative RMS errors of this approximation
against the CFT target are
\begin{equation}
\begin{aligned}
\epsilon_T^{(2\ {\rm vertices})}&=0.699\pm0.072,\\
\epsilon_T^{(3\ {\rm vertices})}&=0.972\pm0.095,\\
\epsilon_T^{(4\ {\rm vertices})}&=0.597\pm0.079.
\end{aligned}
\label{appnum-eq:historical-absolute-ward-errors}
\end{equation}
Errors are standard errors across eight feature seeds. At the same
settings, the geometric mean of the paired RBF-to-Log-Kernel error
ratios is
\begin{equation}
\frac{\epsilon_{T,\rm RBF}}{\epsilon_{T,\rm selected}}
=
31.74,
\qquad
95\%\ \mathrm{CI}=[27.39,36.78],
\label{appnum-eq:ward-rbf-ratio}
\end{equation}
Ratios are formed at matching width, configuration, and feature seed.  The
log ratios are averaged across the three configurations within each seed,
and a Student-$t_7$ interval is formed from the eight independent
seed means. Figure~\ref{appnum-fig:jhep-evidence-ward-independent-blocks}
shows both the absolute errors and the paired comparison. The
relative improvement distinguishes the radial samplers at an
accuracy set by \eqref{appnum-eq:historical-absolute-ward-errors}.

Errors from the explicit network averages scale as $M^{-1/2}$ and
are approximately flat in $N$ over the sampled range. At $N=2000$
and $M=32\,000$, the mean relative RMS errors range from $340$ to
$670$ across the three tapered configurations.
Section~\ref{appnum-app:ward-statistics} specifies the sampling units.

Exact phase integration removes this source of sampling noise. The
characteristic function of the cosine field is a product of Bessel
functions, from which both Ward numerators can be evaluated at fixed
momenta. The formulas in section~\ref{appnum-app:exact-phase-ward} retain
the finite-width phase cumulants. Averaging over independent momentum
blocks with the corresponding characteristic-function weights then
estimates the full ensemble ratios.

For a separate neutral two-vertex benchmark, take charges
$(0.3,-0.3)$ at $w=(-1/2,1/2)$ and the six probes
$z=\pm1/2+iy$, $y=0.2,0.35,0.5$. We use $\tau=0.15$ and spectral
pairs $(\Lambda_{\rm IR},\Lambda_{\rm UV})=(b^{-1},b)$ with
$b=30,100,300$ in the reference units. At $N=2^{21}$, the mean
conditional relative RMS stress-Ward error across eight independent
momentum samples is $0.0651\pm0.0033$ for $b=100$ and
$0.1087\pm0.0103$ for $b=300$. The corresponding Gaussian
finite-regulator biases are $0.05265$ and $0.00624$.
Increasing the spectral window reduces the bias while raising the
frequency-sampling requirement.

\begin{figure}[t]
\centering
\includegraphics[width=0.94\linewidth]{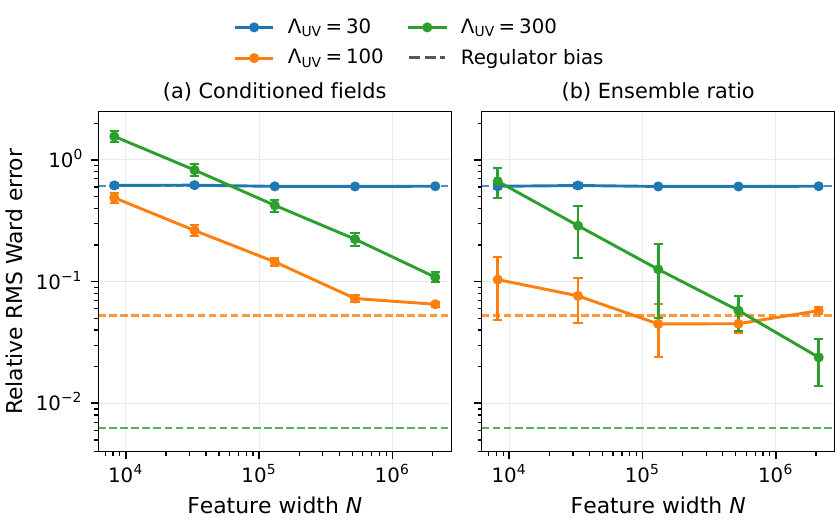}
\caption{Exact phase-integrated stress-Ward errors for the neutral
two-vertex benchmark. Left: mean conditional relative RMS error of
eight independently sampled momentum sets; bars give one standard
error. Right: relative RMS error of the full ensemble ratio estimated
with characteristic-function-weighted block means; bars give a
delete-one-block jackknife standard error of the RMS statistic.
The spectral pairs are $(\Lambda_{\rm IR},\Lambda_{\rm UV})=(b^{-1},b)$
with $\tau=0.15$. Dashed lines give the infinite-width Gaussian
finite-regulator biases. Every curve uses the same six prescribed
probes. Widths share prefixes within each seed; independent seeds
provide the sampling blocks.}
\label{appnum-fig:exact-phase-ward}
\end{figure}

The characteristic-function-weighted ensemble ratios at this width
have relative RMS errors $0.0577$ and $0.0239$ for $b=100$ and $300$,
respectively. Delete-one-block jackknife standard errors of these
RMS statistics are $0.0037$ and $0.0099$. These describe the six-probe
benchmark; the individual conditioned fields and the ensemble ratio
are shown separately in figure~\ref{appnum-fig:exact-phase-ward}. At
$b=300$, the mean difference between the exact conditional stress
ratio and its Gaussian covariance approximation is
$1.10\times10^{-5}$ in the same CFT-normalized norm.

Vertex correlations provide a further charged-sector comparison.  The fitted
weights follow $h_a=a^2/2$ over the studied charges with the smooth
regulator; the sharp regulator gives larger uncertainty at the largest
charge.

\subsection{Fermionic and boundary observables}
\label{appnum-sec:boundary-numerics}

The two fermionic feature constructions share the canonical target but
sample their retained pairings differently.  The coefficient bound in
appendix~\ref{app:pairing-residuals} relates their pairing errors to the
super-Virasoro residuals before canonicalization.
For the boson-derived and independent Cauchy constructions, all six
fitted exponents before canonicalization are compatible with $-1/2$.  For a fixed
finite half-integer mode window define the raw CAR pairing error by
\begin{equation}
\epsilon_{\rm CAR}(N,L)=
\frac{\|\eta_{N,L}-\eta_0\|_F}{\|\eta_0\|_F},
\qquad (\eta_0)_{rs}=\delta_{r+s,0}.
\label{appnum-eq:raw-car-defect}
\end{equation}
The raw $[L,G]$ and $\{G,G\}$ coefficient residuals are evaluated before
canonicalization on the same fixed normal-ordered PBW basis, with the
canonical finite-mode remainder subtracted as in the bosonic calculation.
The limiting canonical boson and fermion modes give the standard
$c=3/2$ super-Virasoro readout on filtered sectors with sufficiently
large oscillator cutoffs. A finite readout uses the projected bosonic
pairing \eqref{eq:fock-pairing-projection} when its mixed block is
positive, together with an admissible retained CAR pairing. The raw
residuals in the figure are computed before this readout. The filtered
domain accounts for modes at the oscillator cutoff.

\begin{figure}[t]
\centering
\includegraphics[width=0.94\linewidth]{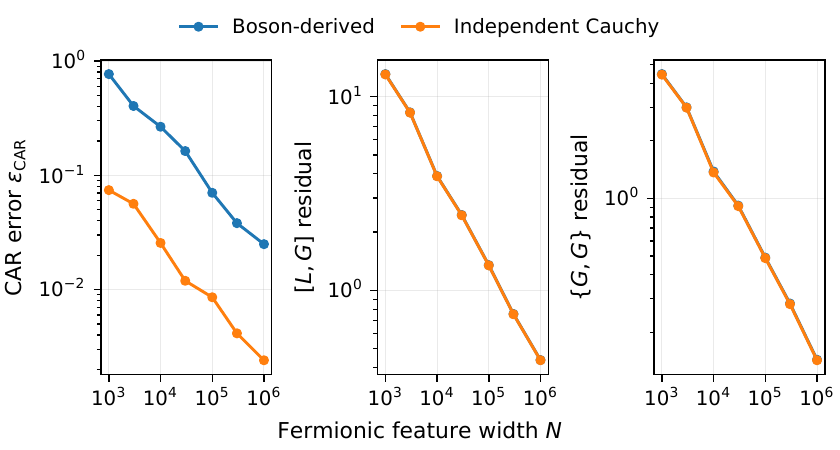}
\caption{Boson-derived and independently sampled Cauchy representations of the fermionic sector. The panels show CAR pairing errors and super-Virasoro residuals before canonicalization.}
\label{appnum-fig:jhep-evidence-ns-two-constructions}
\end{figure}

\begin{figure}[t]
\centering
\includegraphics[width=\textwidth]{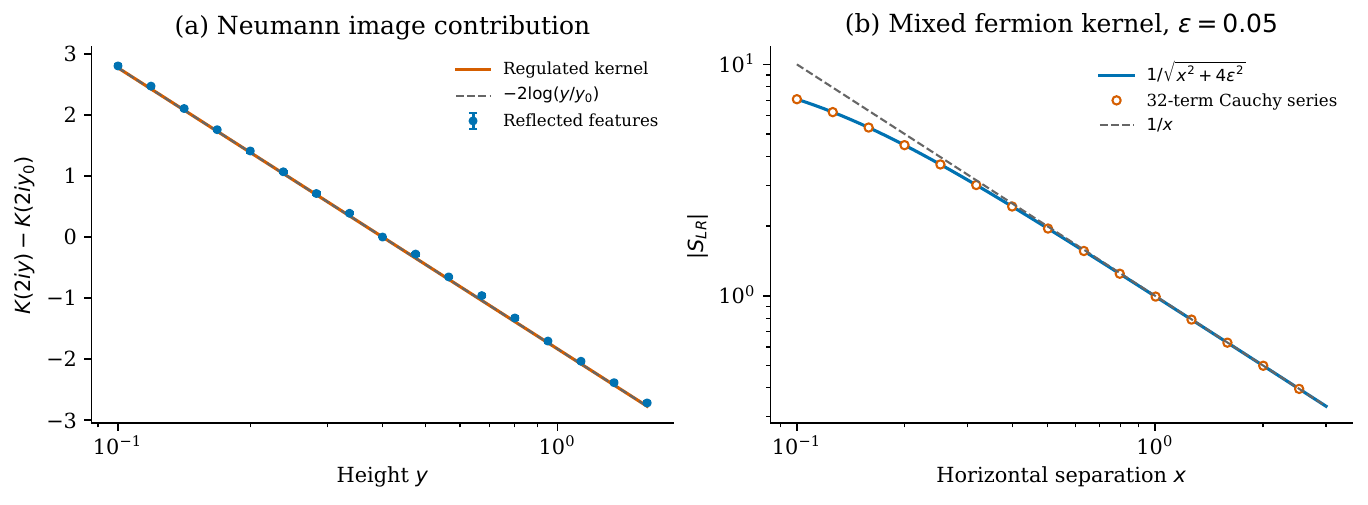}
\caption{Boundary kernels from reflection.  (a) Neumann variance
differences relative to $y_0=0.4$, with the scalar bulk coincident term
removed.  Points are phase-averaged estimates from 32 independent batches
of $16\,384$ momenta; bars give one standard error of the batch mean.
The solid line uses the raised-cosine regulator with
$(\Lambda_{\rm IR},\Lambda_{\rm UV},\tau)=(10^{-3},10^3,0.15)$; the
dashed line is the continuum logarithm.  (b) Magnitude of the mixed
fermion kernel at height $\varepsilon=0.05$.  The markers evaluate a
32-term Cauchy series, the solid line is the exact image kernel, and the
dashed line is its large-separation asymptote.  Numerical methods are
given in section~\ref{appnum-app:boundary-methods}.}
\label{appnum-fig:boundary-kernels}
\end{figure}

Reflection provides a complementary test of the field kernels through
their dependence on distance to the boundary.  For the scalar, subtracting variances at two heights cancels the
bulk coincident term:
\begin{align}
V_\sigma(y;y_0)
&:=\mathbb E[\Phi_{\sigma,N}(iy)^2]
-\mathbb E[\Phi_{\sigma,N}(iy_0)^2]\nonumber\\
&=\sigma\bigl[K_{\Lambda,\tau}(2iy)
-K_{\Lambda,\tau}(2iy_0)\bigr]
\longrightarrow -2\sigma\log(y/y_0).
\label{appnum-eq:boundary-height-statistic}
\end{align}
The logarithmic slope is $-2\sigma$ in the normalization
$K=-\log|z-w|^2$ used throughout this paper.  For Neumann conditions,
figure~\ref{appnum-fig:boundary-kernels}(a) compares this prediction with the
regulated kernel and phase-averaged feature estimates.  The slope from 32
independent frequency batches is $-1.991\pm0.008$ (one standard error),
using $N=16\,384$ features per batch.  The matching finite-cutoff kernel has
slope $-1.999995$ over the same heights.  These error bars describe frequency
sampling after phase averaging.

For the fermion, take $z=x+i\varepsilon$ and $w=i\varepsilon$.  The mixed
correlator is
\begin{equation}
S_{LR}(z,w)=\frac{\eta}{x+2i\varepsilon},\qquad
|S_{LR}(z,w)|=\frac1{\sqrt{x^2+4\varepsilon^2}}.
\label{appnum-eq:boundary-fermion-geometry}
\end{equation}
The height controls the short-distance crossover. For
$|x|\gg2\varepsilon$, the magnitude approaches $1/|x|$.
Figure~\ref{appnum-fig:boundary-kernels}(b) displays this crossover and its
deterministic finite-Cauchy representation.  The 32-term series differs
from the continuum mixed kernel by less than $6.6\times10^{-12}$ in
relative magnitude over the displayed points.  

\section{Numerical methods and statistical analysis}
\label{appnum-app:numerical-methods}
\subsection{Sampling and contour geometry}
\label{appnum-app:controls}
\label{appnum-app:cauchy-sampling}

Spectral cutoffs $(\Lambda_{\rm IR},\Lambda_{\rm UV})$, oscillator cutoff
$L$, feature width $N$, contour geometry, and the number $M$ of sampled
networks are varied independently.  Regulator dependence is reported
separately from sampling uncertainty.  Parameter settings, seeds, individual
fits, and exclusions are retained in the accompanying computational archive.

For the radius comparison, adjacent points of the $(R_0,\rho)$ grid are
declared compatible when their pairing difference is bounded by the sum of
the Cauchy--Green envelope and a $3\sigma$ finite-feature term.  All ten $R_0$
nodes belong to the connected region for the tapered regulator.  The median
ratio of the measured variation to the analytic envelope is $0.017$, with
maximum $7.4$; the finite-feature uncertainty term is therefore needed on
some edges.  The large non-holomorphy of the sharp regulator gives a permissive
Cauchy--Green bound. We therefore assess radius variation together with
the current-kernel error.

The architecture comparisons vary an RBF radial prior, a local radial-exponent direction,
the signs and mixing of mode pairings, angular isotropy, and phase
correlations.  Field symmetries are evaluated on the two-argument kernel
before contour projection.  In particular, translation invariance compares
$K(x+a,y+a)$ with $K(x,y)$ using the actual feature-index correlations of
each architecture.  Rotational defects are evaluated independently of the
contour-pairing error.

The local exponent direction is
\begin{equation}
p_\varepsilon(k)
=
\frac{1}{k^2}
\left[1+\varepsilon\bigl(\log k-\langle\log k\rangle\bigr)\right],
\end{equation}
with positivity enforced on the annulus.  The mode-pairing, angular, and phase changes are evaluated using the same two-argument kernel and intrinsic pairing diagnostics as in section~\ref{appnum-sec:audits}.

The discrete Cauchy features introduce a separate sampling law over
mode indices.  For a geometric proposal $q_n=(1-q)q^n$ on contour ratio
$r=|w/z|<1$, finite second moment requires
\begin{equation}
{q>r^2.}
\label{appnum-eq:cauchy-proposal-condition}
\end{equation}
Indeed
$\mathbb E_n|z^{-n-1}w^n/q_n|^2
=|z|^{-2}\sum_{n\ge0}r^{2n}/q_n$,
which is finite precisely when $q>r^2$ for the infinite-support proposal.
The numerical comparison uses $q=0.75$ and the renormalized proposal on
$0\le n\le400$.  It is unbiased for the truncated geometric kernel, whose
relative tail error on the nested contours is $r^{401}$; the proposal mass
omitted is $q^{401}$.  Every fixed truncation has finite variance, including
below the infinite-support threshold.  The divergence is measured as
that truncation increases.  On a fixed finite mode window inside this
support, contour extraction has no tail bias.  If $\widehat p_N(n)$ is the
empirical mode frequency and $q_n$ the actual proposal probability, then
\begin{equation}
\eta_{rs}^{(N)}=\delta_{r+s,0}
\frac{\widehat p_N(|r|-1/2)}{q_{|r|-1/2}}.
\label{appnum-eq:cauchy-empirical-pairing}
\end{equation}
Each selected amplitude has mean one and variance $(1-q_n)/(Nq_n)$.
A finite-window canonical CAR readout additionally requires every retained
mode to have been sampled: a missing mode gives a zero pairing amplitude and makes the retained
pairing degenerate.  For a fixed retained
index set $I$, the failure probability is at most
$\sum_{n\in I}(1-q_n)^N$.  The native boson-derived readout likewise requires
an admissible fermionic pairing in addition to the bosonic Fock
criterion.

\subsection{Composite observables and correlated uncertainties}
\label{appnum-app:finite-width}
\label{appnum-app:ward-statistics}
\label{appnum-app:tt-statistics}
\label{appnum-app:width-fits}
\label{appnum-app:exact-phase-ward}
\label{appnum-app:four-point-width}

For each observable we specify whether the features have been averaged
analytically or retained as a fixed sample. Explicit network averages
are evaluated in independent sampling blocks.

Covariance normal ordering uses the same conditioned feature set as the
observable.  Thus
\begin{equation}
T_N(z)
=\frac12\left[J_N(z)^2-K_{JJ}^{(N)}(z,z)\right],
\end{equation}
and
\begin{equation}
V_{a,N}(w,\bar w)
=\exp\left[
ia\phi_N(w,\bar w)
+\frac{a^2}{2}K_{\phi\phi}^{(N)}(w,w)
\right].
\end{equation}
The infrared-normalized continuum vertex is defined in
\eqref{eq:infrared-renormalized-vertex}. Its multiplicative
normalization cancels between the numerator and denominator of
each Ward ratio, so the finite-cutoff comparisons here are unchanged.

The self-contractions are conditional expectations given the
momenta. With $c_z(\mathbf k)=
\mathbb E_\gamma[J_N(z)^2\mid\mathbf k]$, the unconditional
connected four-current cumulant gives the general relation
\begin{equation}
G_{TT}^{\rm dir}-\frac12K_{JJ}(z,w)^2
=\frac14\kappa_{JJJJ}(z,z,w,w)
-\frac14\operatorname{Cov}_{\mathbf k}(c_z,c_w).
\label{appnum-eq:conditional-tt-cumulant}
\end{equation}
Here covariance is bilinear, without complex conjugation. It follows
by first expanding
$\mathbb E[(J_N(z)^2-c_z)(J_N(w)^2-c_w)]$ at fixed
momenta. For the independent-phase cosine architecture,
$c_z=-\sigma_\phi^2(8N)^{-1}\sum_j\overline{k_j}^{\,2}$
is independent of $z$. Isotropy makes both its mean and its bilinear
second moment vanish, giving \eqref{appnum-eq:direct-wick-cumulant}.
The separated derivative cumulant is retained explicitly in this
calculation.

For exact phase integration, write $s_N=\sigma_\phi/\sqrt N$ and
$\theta_j(x)=\operatorname{Re}(\overline{k_j}x)$. A neutral vertex
source factorizes through
\begin{equation}
Q_j=\sum_i a_i e^{i\theta_j(w_i)},\qquad
t_j=s_N|Q_j|,\qquad
Z_N(\mathbf k):=
\mathbb E_\gamma\!\left[e^{i\sum_i a_i\phi_N(w_i)}
\mid\mathbf k\right]
=\prod_jJ_0(t_j).
\label{appnum-eq:phase-characteristic-function}
\end{equation}
The vertex normal-ordering factors cancel in the ratios. At fixed
spectral cutoff they are independent of the momenta, since the
conditional scalar variance is $\sigma_\phi^2/2$ at every point.

Define $\delta_j(z)=\theta_j(z)-\arg Q_j$ and
\begin{equation}
\mu_j(z)=\frac{s_N\overline{k_j}}2\,
\sin\delta_j(z)\,\frac{J_1(t_j)}{J_0(t_j)}.
\end{equation}
For $Z_N(\mathbf k)\ne0$, exact phase integration gives
\begin{align}
R_{J,N}^{\rm phase}(z\mid\mathbf k)
&=\sum_j\mu_j(z),\nonumber\\
R_{T,N}^{\rm phase}(z\mid\mathbf k)
&=\frac12\left[\sum_j\mu_j(z)\right]^2
-\frac12\sum_j\mu_j(z)^2\nonumber\\
&\quad-\frac{s_N^2}{16}
\sum_j\overline{k_j}^{\,2}\cos\!\bigl(2\delta_j(z)\bigr)
\frac{J_2(t_j)}{J_0(t_j)}.
\label{appnum-eq:exact-phase-ward}
\end{align}
The terms with $Q_j=0$ are defined by continuity. To obtain these
expressions, shift each phase by $\arg Q_j$ and use
$\langle e^{it\cos\gamma}\rangle=J_0(t)$,
$\langle\cos\gamma\,e^{it\cos\gamma}\rangle=iJ_1(t)$
and $\langle\cos2\gamma\,e^{it\cos\gamma}\rangle=-J_2(t)$.
The current contribution of feature $j$ is
$-is_N\overline{k_j}\sin(\theta_j+\gamma_j)/2$.
Separating its same-feature second moments from its cross-feature
products gives the two correction terms in the stress ratio.

The expansions $J_1(t)/J_0(t)=t/2+O(t^3)$ and
$J_2(t)/J_0(t)=t^2/8+O(t^4)$ recover
\eqref{appnum-eq:gaussian-ward-proxy}. At fixed cutoffs, charges and
insertion points, their corrections are $O(N^{-1})$. The constants
depend on the momentum moments, which must also be controlled when
the spectral window grows.

A conditional ratio must be weighted by its vertex characteristic
function when the momenta are averaged. The full ensemble ratio and
its estimator from $B$ independent momentum samples are
\begin{equation}
R_{T,N}^{\rm ens}
=\frac{\mathbb E_{\mathbf k}[Z_N R_{T,N}^{\rm phase}]}
{\mathbb E_{\mathbf k}[Z_N]},\qquad
\widehat R_{T,N}^{\rm ens}
=\frac{\sum_{b=1}^B Z_{N,b} R_{T,N,b}^{\rm phase}}
{\sum_{b=1}^B Z_{N,b}}.
\label{appnum-eq:phase-weighted-ward-estimator}
\end{equation}
The numerator and denominator use the same blocks. The benchmark has
$B=8$, positive $J_0(t_j)$ factors, and computes $Z_N$ through the sum
of their logarithms.

The width grid is $N=2^{13},2^{15},2^{17},2^{19},2^{21}$.
For each spectral pair and seed, the widths use prefixes of one
maximum-width independent momentum sample. Different seeds and
spectral pairs are independent. Error bars for the mean conditional
relative RMS error are standard errors over the eight seeds.
For the RMS error of \eqref{appnum-eq:phase-weighted-ward-estimator}, we
delete each momentum block together with its numerator and denominator
and report the jackknife standard error of that statistic. These
summaries describe the prescribed six probes and width grid.

The infinite-width Gaussian reference at the same regulator is
\begin{equation}
R_J^{\Lambda,\tau}(z)
=\sum_i a_i\frac{\overline{z-w_i}}{|z-w_i|}
\int_{\Lambda_{\rm IR}}^{\Lambda_{\rm UV}}
W_\tau(u(k))J_1(k|z-w_i|)\,dk,\qquad
R_T^{\Lambda,\tau}(z)=\frac12[R_J^{\Lambda,\tau}(z)]^2.
\label{appnum-eq:regulated-ward-reference}
\end{equation}
Quadrature in log radius is split at the taper joins and resolves
the Bessel oscillations. Its relative RMS biases against the CFT
target are $0.60778$, $0.05265$ and $0.00624$ for
$b=30,100,300$. The same CFT norm is used for the sampled errors,
the differences from the regulated reference, and the difference
between exact phase integration and the Gaussian proxy. Direct
two-feature phase quadrature checks both numerators and the
characteristic function to $2\times10^{-15}$ in absolute error.
The repository retains every seed, probe and regulator value.

For Ward ratios, numerator and denominator means use the same network
realizations.  Their covariance is retained within each block.  The
dependence on the number $M$ of network realizations is fitted separately
from the dependence on feature width $N$.

The architecture comparison applies the uniformly weighted
relative RMS error \eqref{appnum-eq:ward-rel-rms} to the Gaussian covariance
proxy \eqref{appnum-eq:gaussian-ward-proxy} on each probe configuration.  At a
fixed width and regulator, the error of a control is divided by that of the
Log-Kernel architecture at the same configuration and seed.  Log ratios are
averaged across configurations within each seed, and the mean and standard
error are then evaluated across the eight independent seeds.
Exponentiating the mean and its Student-$t_7$ interval gives the
geometric ratio and 95\% interval in \eqref{appnum-eq:ward-rbf-ratio}.
The tapered
comparison contains three configurations per seed, or twenty-four paired
measurements; the seed is the independent sampling unit.

Each stress-tensor block supplies a four-component vector of ratios to the
matched analytic regulator target.  Let $\bar q$ denote the mean of the
twelve block vectors and $S$ their sample covariance.  The GLS estimate in
\eqref{appnum-eq:tt-hardening-result} uses this covariance, and its quoted
uncertainty is a plug-in standard error.  The four components within a block
are correlated.  Table~\ref{appnum-tab:tt-geometry} gives their marginal means and
standard errors.

\begin{table}[htbp]
\centering
\begin{tabular}{@{}rcc@{}}
\toprule
Separation & Direct $TT$ ratio & Wick ratio \\
\midrule
0.3106 & $0.9918\pm0.0350$ & $1.00145\pm0.00161$ \\
0.3857 & $0.9786\pm0.1380$ & $0.99745\pm0.00282$ \\
0.4592 & $1.0034\pm0.2542$ & $0.99233\pm0.00389$ \\
0.5307 & $1.9205\pm0.5621$ & $0.99412\pm0.00595$ \\
\bottomrule
\end{tabular}
\caption{Matched finite-regulator $TT$ ratios.  Errors are standard errors
across twelve independent blocks; geometries within a block are correlated.
The covariance tests in section~\ref{appnum-app:tt-statistics} use all four geometries.}
\label{appnum-tab:tt-geometry}
\end{table}

The joint Hotelling tests compare the four-dimensional mean with
$(1,1,1,1)$, giving $p=0.246$ (direct) and $p=0.319$ (Wick).  To examine
variation about an unspecified common ratio, choose three linearly
independent geometry contrasts with $C\mathbf 1=0$.  The statistic and its
Gaussian-block reference distribution are
\begin{equation}
Q=12(C\bar q)^T(CSC^T)^{-1}(C\bar q),\qquad
\frac{9}{33}Q\sim F_{3,9}.
\label{appnum-eq:tt-contrast-calibration}
\end{equation}
Here $S$ is the covariance of individual block vectors.  For the direct estimator, $Q=9.185$ gives $F=2.505$ and $p=0.125$;
the Wick contrast test gives $p=0.180$.  This finite-sample calibration is
exact for identically distributed multivariate Gaussian blocks and
approximate for the Monte Carlo block averages \cite{NIST:Hotelling}.  All
four geometries and all twelve blocks enter both comparisons.  These tests concern the joint mean and the chosen geometry contrasts.
Pointwise precision varies across the geometries, as shown in the table.

The dependence on width is summarized across eight observables with
an expected $N^{-1/2}$ random-feature sampling law.
Their inverse-variance pooled fit is
\begin{equation}
b_{\rm pooled}=-0.499102\pm0.004661,
\qquad
Q=0.775\ \text{on }7\ \text{dof},
\qquad
p=0.998,
\qquad
I^2=0\%.
\label{appnum-eq:pooled-random-feature-rate}
\end{equation}
Ward errors, exact stationarity, and angular controls with a persistent
defect have different null behavior and are excluded from this pool.
Shared feature samples can correlate the fitted exponents. We report
Cochran's $Q$ and $I^2$ as descriptive summaries of this pool.

A separate equal-weight average of the eight exponents gives $-0.4981$.
We resample the control and NS seed families independently, using the
same draw across observables and widths within each family, and refit
the logarithms of the mean errors. With $1000$ replicates and random
seed 20261118, the 95\% percentile interval is $[-0.5331,-0.4663]$.
This average differs from inverse-variance pooling. Convergence is
supported by the sampling estimate \eqref{eq:two-error-pairing-bound}
and the separate width fits; the resampling describes their joint variation.

The scalar four-point relation \eqref{eq:four-point-width-scaling}
follows by grouping the feature indices in the product of four
network sums. There are $N$ terms with a single feature index
and $N(N-1)$ terms for each of the three pairings, so
\begin{equation}
G_{4,N}
=\frac{N(N-1)}{N^2}W_4+\frac{N}{N^2}\E[X_1X_2X_3X_4].
\label{appnum-eq:four-point-feature-counting}
\end{equation}
Writing the normalized spectral characteristic function as
$F_{\Lambda,\tau}(v)=\int d^2k\,p_{\Lambda,\tau}(k)\cos(k\cdot v)$,
uniform phase averaging gives
\begin{equation}
\begin{aligned}
\E[X_1X_2X_3X_4]=\frac{\sigma_\phi^4}{8}\bigl[
&F_{\Lambda,\tau}(x_1+x_2-x_3-x_4)\\
+&F_{\Lambda,\tau}(x_1+x_3-x_2-x_4)\\
+&F_{\Lambda,\tau}(x_1+x_4-x_2-x_3)\bigr].
\end{aligned}
\label{appnum-eq:four-point-single-feature}
\end{equation}
Together with $K_{ij}=\sigma_\phi^2F_{\Lambda,\tau}(x_i-x_j)/2$,
this specifies the connected coefficient for arbitrary
insertion points in the full momentum-and-phase ensemble.

At coincident points,
$\langle\cos^2\gamma\rangle=1/2$ and
$\langle\cos^4\gamma\rangle=3/8$, giving
\begin{equation}
\kappa_4(N)=-\frac{3\sigma_\phi^4}{8N},
\end{equation}
and standardized excess kurtosis $-3/(2N)$. The saved
calculation uses eight independent blocks at each width
$N=8,16,32,64,128$, with $10^6$ network realizations per
block. A least-squares fit of $\log|\kappa_4|$ to $\log N$,
using the block-mean cumulant at each width, gives the
tapered-window exponent $-0.9951\pm0.0092$ quoted in the
main text; the uncertainty is the regression standard error.
The exponent and its $R^2=0.9997$ are also reported in
appendix~\ref{appnum-sec:composite-checks}. The cumulant
quantifies the finite-width departure from the Gaussian
limit.

\subsection{Boundary kernel calculations}
\label{appnum-app:boundary-methods}

The scalar calculation uses the same raised-cosine probability law and
normalization as section~\ref{sec:lk}.  At fixed sampled momenta, averaging
over the uniform phases gives
\begin{equation}
\widehat K_N(a,b)=\frac{2L_w}{N}\sum_{j=1}^N
\cos\bigl(k_j\cdot(a-b)\bigr).
\end{equation}
The reflected conditional covariance is the four-term expression
\begin{equation}
\widehat K_N^\sigma(z,w)=\tfrac12\bigl[
\widehat K_N(z,w)+\sigma\widehat K_N(z,\bar w)
+\sigma\widehat K_N(\bar z,w)+\widehat K_N(\bar z,\bar w)\bigr].
\label{appnum-eq:boundary-four-term-covariance}
\end{equation}
On the diagonal, its difference between heights reduces exactly to
\begin{equation}
\widehat V_\sigma(y;y_0)=\frac{2\sigma L_w}{N}\sum_{j=1}^N
\bigl[\cos(2k_j^y y)-\cos(2k_j^y y_0)\bigr].
\label{appnum-eq:boundary-feature-estimator}
\end{equation}
This estimator averages the phases analytically and samples the
frequency-dependent conditional covariance.  Radial momenta are drawn by rejection
sampling from $W_\tau(u)$ in log radius and angles are uniform.  We use
$N=16\,384$, 32 independent batches, seed 20261107, 17 logarithmically
spaced heights $0.1\le y\le1.6$, and $y_0=0.4$.  Each batch uses common
momenta at all heights.  A straight-line fit with an intercept in
$\log(y/y_0)$ is performed in each batch; the standard error quoted in
section~\ref{appnum-sec:boundary-numerics} is the sample standard deviation of those 32
slopes divided by $\sqrt{32}$.  Correlations among the heights are thereby
retained.

The regulated reference is
\begin{equation}
V_\sigma^{\rm reg}(y;y_0)=2\sigma\int_{\Lambda_{\rm IR}}^{\Lambda_{\rm UV}}
\frac{dk}{k}\,W_\tau(u(k))
\bigl[J_0(2ky)-J_0(2ky_0)\bigr].
\end{equation}
We split the quadrature at the taper joins and into intervals resolving
the Bessel oscillations.  Tightening its tolerance and halving the maximum
interval length checks the numerical reference.  The largest deviation of
the regulated reference from $-2\log(y/y_0)$ is $2.8\times10^{-5}$
over these heights.

The fermion calculation uses the deterministic truncation already defined
in section~\ref{sec:fermion-architectures},
\begin{equation}
C_L^>(u,v)=\frac{1-(v/u)^L}{u-v},\qquad |v|<|u|.
\end{equation}
For the mixed entry, $u=x+i\varepsilon$ and $v=-i\varepsilon$, so
the relative truncation error is $|v/u|^L$.  We take $L=32$,
$\varepsilon=0.05$, and 60 logarithmically spaced points
$0.1\le x\le3$.  The largest error is bounded by
$(1/\sqrt5)^{32}=6.5536\times10^{-12}$.
The exchange of the two ordered arguments includes the fermion minus
sign.  The two reflected evaluations share their mode features.

\section{Finite-mode and readout checks}
\label{appnum-app:formal-audit}

\subsection{Pairing and projection checks}

For each architecture and width we evaluate the pairing $\Omega$ before
canonicalization and compute
\begin{equation}
\delta_\star,
\qquad
\delta_{\rm iso},
\qquad
\delta_H,
\qquad
\lambda_{\min}\!\left[D^{-1/2}\frac{G+G^\dagger}{2}D^{-1/2}\right],
\qquad
\delta_G
=\frac{\|D^{-1/2}GD^{-1/2}-I\|_F}{\sqrt L}.
\end{equation}
The raw matrices are first tested against
\eqref{eq:physical-darboux-criterion}. We then apply the projection
\eqref{eq:fock-pairing-projection} and test the eigenvalues of $H_N$.
For a positive $H_N$, both the principal canonicalizer with $G=H_N$
and \eqref{eq:normalized-fock-canonicalizer} are checked against the
canonical form and the adjoint involution. The raw matrix, projection
correction and signed eigenvalues are retained separately.

We evaluate adjoint compatibility on $\Omega$ itself. A numerical
Darboux map has a coordinate-dependent adjoint defect; $\delta_G$
instead measures the cross-block amplitude relative to the canonical
target, independently of the positive-frequency block.

\subsection{Filtered-sector and coefficient checks}

Alongside these polarization checks, we verify the finite-level support
condition directly.  Equation~\eqref{eq:filtered-virasoro} gives the
sufficient support bound.
For each mode pair we evaluate
\begin{equation}
K_{\max}(L;m,n)=L-m_--n_-
\end{equation}
whenever the right-hand side is nonnegative and verify the identity on a
sparse Fock basis for several charged sectors.  Outside the guaranteed domain,
the truncated commutator contains an edge-supported oscillator polynomial.
We retain this cutoff contribution explicitly in the commutator remainder.

For the commutator estimate, the normal-ordered
Poincar\'e--Birkhoff--Witt (PBW) basis includes $\alpha_0$ and is held fixed as
$\Omega$ varies.  We compute both the canonical finite-mode remainder
$\mathcal R(\Omega_0)$ and its pairing-dependent excess
$\Delta\mathcal R(\Omega)=\mathcal R(\Omega)-\mathcal R(\Omega_0)$.  The
five-decade synthetic $\varepsilon$-sweep along the reference perturbation
direction gives a bounded stability ratio $1.6502$--$1.8151$; twelve
independent perturbation directions give $1.5583$--$2.1106$.  The synthetic sweep holds the mode window fixed. The width-grid
calculation samples perturbations induced by the features.

The canonical finite-mode central term agrees with the standard one on
the filtered domain.  Additional commutator terms outside that domain depend
on the oscillator cutoff and are included in the explicit remainder.

\bibliographystyle{JHEP}
\bibliography{Virasoro_JHEP_2026}

\end{document}